\documentclass[twocolumn,preprintnumbers,amsmath,amssymb,pra]{revtex4}

\usepackage{graphicx,amsmath}
\usepackage{dcolumn}
\usepackage{bm}
\usepackage{amssymb}
\usepackage{epstopdf}
\usepackage{color}
\graphicspath{{Figures/}}

\begin{document}

 \title{Quantum correlations of the photon fields in a waveguide quantum electrodynamics}

\begin{abstract}
We present a time-dependent quantum calculations of the first-order and second-order photon correlation functions for the
scattering of a single-photon pulse on a two-level atom (qubit)
embedded in a one-dimensional open waveguide. Within Markov
approximation we find the analytic expression for the quantum
operator of positive frequency electric field. We restricted
Hilbert space of initial states by the states with one and two
excitations and show that the photon probability amplitudes are
given by the off-diagonal matrix elements of the electric field
operator between these states. For two-excitation initial state
where the atom is excited and there exists a single photon in a
waveguide we calculate the second-order correlation function which
describes the measurements by two detectors at two different
space-time points. The second-order correlation function exhibits
the interference term showing that the measurements of two
detectors are correlated. This interference is similar to that
found in the Hanbury Brown and Twiss correlation experiment with
two indistinguishable photons.

\end{abstract}


\date{\today}

\author{Ya. S. Greenberg}\email{yakovgreenberg@yahoo.com}
\affiliation{Novosibirsk State Technical University, Novosibirsk,
Russia}
\author{O. A. Chuikin} \affiliation{Novosibirsk State
Technical University, Novosibirsk, Russia}

\author{A. G. Moiseev} \affiliation{Novosibirsk State
Technical University, Novosibirsk, Russia}

\author{A. A. Shtygashev} \affiliation{Novosibirsk State
Technical University, Novosibirsk, Russia}

\maketitle

\section{Introduction}

Theoretical and experimental studies of the interaction between
quantum emitters and photons in confined geometries underpin
modern technologies aimed at the development of quantum processors
and quantum information processing circuits. Recently, significant
progress has been made in one-dimensional quantum electrodynamics
technologies, namely, waveguide quantum electrodynamics (wQED)
\cite{Rai2001, Roy17, Chang18, Turschmann19, Sheremet23, Gu2017}
that enables atoms to couple directly to photon travelling modes.

A single photon scattering through an array of two-level atoms
embedded in a 1D open waveguide has been extensively studied both
theoretically \cite{Ruos2017,Lal2013,Chang2012} and experimentally
\cite{Mirho2019,Brehm2021,Loo2014}. There are many theoretical
approaches describing the photon scattering from the atoms
embedded in 1D open waveguide. These studies have been performed
both within the framework of stationary scenario
\cite{Shen2009,Cheng2017,Fang2014,Zheng2013, Roy2011, Huang2013,
Diaz2015, Fan2010, Lal2013, Kii2019, Green2015,
Green2021,Tsoi2008} and time-dependent dynamical methods
\cite{Chen2011, Liao2015, Liao2016a, Liao2016b, Zhou2022,
Green2023a, Green2023b,Green2014}.

In the present paper we consider the scattering of a single photon
pulse from a two-level atom embedded in a one-dimensional open
waveguide. However, as distinct from the other methods, here, we
consider the waveguide electric field as an explicit dynamic
component of our model. The derivation of the explicit expression
for the quantum operator of the positive frequency electric field
is based on the Heisenberg equations for the field and atomic
operators which are transformed into a simplified form by means of
a Markov approximation. This approach allows us to calculate the
photon amplitudes as the off-diagonal matrix elements of the
quantum operator of the positive frequency electric field. Using
these matrix elements we can construct the first and second-order
photon correlation functions.

The first-order correlation function provides the probability
density for the scattered field measured by a single detector at a
given space-time point $x,t$. Its Fourier transform gives the
transmitted and reflected photon spectra.

The second-order correlation function describes the measurements
by two detectors at two different space-time points, $x_1,t_1$ and
$x_2,t_2$. As distinct from the free fields, here the operators of
the positive frequency fields do not commute at different
space-time points . It means that the measurement by the first
detector influences the later measurements by the second detector.
In other words, the measurement by the first detector changes the
state of the remaining electric field which is later measured by
the second detector. Therefore, the outcomes of the measurements
of two detectors which are described by the second-order
correlation function should inevitably be correlated. This
correlation appears as the interference term in the second-order
correlation function. This interference is similar (but not
identical) to the Hanbury Brown and Twiss (HBT) effect observed in
their correlation experiments with two indistinguishable photons
\cite{Brown1956}.

With regard to photon statistics in one-dimensional waveguide systems with embedded qubits the second-order correlation function was studied in \cite{Zhang2018} by using a quantum jump approach. The construction of higher order correlation functions was developed in \cite{Lu2023} within the approach based on the $n$-photon scattering matrix ($S$-matrix).

Our approach here is similar to the one described in
\cite{Dom2002} where the Poynting vector was calculated as the
average of electric field operators over the initial state with
the atom being initially in the ground state. In order to
concentrate our  study on the second-order correlation function,
here we consider mainly the scattering of a single photon Fock
state light pulse from the excited atom. Our treatment is valid
for any shape of the light pulse. However, in order to obtain
simple and transparent results the specific calculations have been
performed for  incident plane wave.

The paper is organized as follows. In Sec. II we present the model
and introduce the basic parameter describing the coupling of the
atomic dipole to a one-dimensional waveguide. In Sec. III we derive
within a Markov approximation, the expression for the electric
field operator which accounts for the interaction of electric
field with a qubit. In Sec. IV we define the first and second-order photon correlation functions for the initial states with
single and two excitations. In Sec. V we perform the calculation of
the matrix elements of the electric field operator for the
single-excitation initial states. In this case, there exists only
first-order correlation function which describes the measurements
by a single detector. In Sec. VI we calculate the matrix elements
of the electric field operator for two-excitation initial state.
In this case, there exist both first order and second-order
correlation functions. The Hanbury Brown and Twiss effect for the
radiation from a two-level atom is considered in Sec. VII.

\section{The model}

We start from the Hamiltonian $H=H_0+H_{JC}$ where:
   \begin{subequations} \label{1}
 \begin{align}
H_0  = \frac{\hbar\Omega}{2} & \left( {1 + \sigma_z } \right)  +
\int\limits_0^\infty \hbar \omega  \left( {a^\dag  (\omega )
a(\omega ) + b^\dag  (\omega ) b(\omega )} \right)d\omega,
\label{1a}
\\
& \begin{gathered} \label{1b} H_{JC}  = \int\limits_0^\infty \hbar
{g\left( \omega  \right)\left( {a^\dag  (\omega )\sigma_-   +
\sigma_+  a(\omega )} \right)d\omega }
  \\
+ \int\limits_0^\infty \hbar {g\left( \omega  \right)\left(
{b^\dag (\omega )\sigma_ -  + \sigma_ + b(\omega )} \right)d\omega
},
\end{gathered}
 \end{align}
   \end{subequations}
where $H_{JC}$ is Jaynes-Cummings Hamiltonian which describes the
interaction of the photon field with a qubit located at the point
$x=0$. $\sigma_z$ is a Pauli spin operator,
$\sigma_z=|e\rangle\langle e|-|g\rangle\langle g|$, where
$|e\rangle$ and $|g\rangle$ are excited and ground states of a
two-level atom. $\Omega$ is a resonant frequency of a qubit,
$\omega$ is a photon frequency, $\sigma_- = {\left|g\right\rangle}
\left\langle e \right|$ and $\sigma _ +  = {\left| e \right\rangle
} \left\langle g \right|$ are the lowering and raising atomic
operators which lower or raise a state of a qubit.

The quantity $g(\omega)$ in (\ref{1b}) is the coupling between
qubit and the photon field in a waveguide \cite{Dom2002}:
   \begin{equation} \label{2}
g(\omega ) = \left( {\frac{\omega } {{4\pi \varepsilon _0 \hbar
v_g A}}} \right)^{1/2} d,
   \end{equation}
where $d$ is the off-diagonal matrix element of a dipole operator,
$A$ is the effective transverse cross section of the modes in
one-dimensional waveguide, $v_g$ is the group velocity of
electromagnetic waves.

We assume that the coupling is symmetrical, i.e.,
it is the same for forward and backward waves. However, some types
of waveguides can have different coupling for forward and backward radiation,
usually called chiral or non-reciprocal
\cite{Lodahl17, RosarioHamman18, Guimond20, Sheremet23}.
This asymmetry can be taken into account by having two different coupling
constants for forward and backward directions,
$g_\rightarrow (\omega)$ and $g_\leftarrow (\omega)$, which leads
to the major differences in photon transmission and reflection
regarding the direction in which the wave is moving. However,
in this work we consider a different problem, so we left the
non-symmetrical waveguides to possible future studies and focus on
symmetrical ones.

Note that the dimension of the coupling constant $g(\omega)$ is
not a frequency, $\omega$, but a square root of frequency,
$\sqrt{\omega}$, and, as it follows from below (\ref{comut}), the
dimension of creation and destruction operators is
$1/\sqrt{\omega}$.

The photon creation and annihilation operators $a^\dag(\omega)$,
$a(\omega)$, and $b^\dag(\omega)$, $b(\omega)$ describe forward
and backward scattering waves, respectively. They are independent
of each other and satisfy the usual continuous-mode commutation
relations \cite{Blow1990}:
  \begin{equation}\label{comut}
\left[ {a(\omega ),a^\dag  (\omega ')} \right] = \left[ {b(\omega
),b^\dag (\omega ')} \right] = \delta(\omega - \omega ').
  \end{equation}

\section{Calculation of the electric field operator}

Quantum operator for positive frequency electric field reads
\cite{Dom2002}:
   \begin{equation} \label{4}
   \begin{gathered}
E^+  (x,t) = i\frac{\hbar } {d}\int\limits_0^\infty  {d\omega }
g(\omega )a(\omega ,t)e^{ikx}  \\
+ i\frac{\hbar }
{d}\int\limits_0^\infty  {d\omega } g(\omega )b(\omega ,t)e^{ -
ikx},
\end{gathered}
   \end{equation}
where $k=\omega/v_g$. For negative frequency field one should take
hermitian conjugate of (\ref{4}), $E^-(x,t)= \left( E^+(x,t)
\right)^\dagger$.

The formal solution of Heisenberg equation of motion for photon
operators $a(\omega,t)$ and $b(\omega,t)$ is as follows:
   \begin{equation} \label{a}
a(\omega ,t){\text{ }} = a(\omega ,0)e^{ - i\omega t}  -
i\int\limits_0^t {g(\omega )e^{ - i\omega (t - \tau )} \sigma _ -
(\tau )d\tau } ,
   \end{equation}
   \begin{equation}\label{b}
b(\omega ,t){\text{ }} = b(\omega ,0)e^{ - i\omega t}  -
i\int\limits_0^t {g(\omega )e^{ - i\omega (t - \tau )} \sigma _ -
(\tau )d\tau } ,
   \end{equation}
where $a(\omega,0)$ and $b(\omega,0)$ are the photon operators
which would describe a free propagation of the photon waves in a
bare waveguide.

Substituting (\ref{a}) and (\ref{b}) to (\ref{4}) we obtain:
   \begin{equation}\label{7}
\begin{gathered}
  E^+  (x,t) = E_{free}^+ (x,t)
\\
+ \frac{\hbar }
{d}\int\limits_0^\infty  {d\omega } g^2 (\omega )\int\limits_0^t
{d\tau e^{i\omega \tau } \sigma_-  (\tau )} e^{ - i\omega \left(
{t - \frac{x}
{v_g }} \right)}
\\
   + \frac{\hbar }
{d}\int\limits_0^\infty  {d\omega } g^2 (\omega )\int\limits_0^t
{d\tau e^{i\omega \tau } \sigma_-  (\tau )} e^{ - i\omega \left(
{t + \frac{x}
{v_g }} \right)},
\end{gathered}
   \end{equation}
where
   \begin{equation}\label{8}
\begin{gathered}
  E_{free}^ +  (x,t) = i\frac{\hbar }
{d}\int\limits_0^\infty  {d\omega } g(\omega )a(\omega ,0)e^{ -
i\omega \left( {t - \frac{x}
{v_g }} \right)}
\\
   + i\frac{\hbar }
{d}\int\limits_0^\infty  {d\omega } g(\omega )b(\omega ,0)e^{ -
i\omega \left( {t + \frac{x}
{v_g }} \right)}.
\end{gathered}
   \end{equation}

According to Markov approximation \cite{Gardiner85, Shen2009}, we
assume that the coupling $g(\omega)$ is a slowly varying function
around the qubit frequency $\Omega$ and makes a major contribution
at resonance with a qubit, $g(\omega) \simeq g(\Omega)$. This is
so called resonance approximation \cite{Mollow1975, Blow1990}
which can be justified if the bandwidth of incident photon source
is much smaller than that of any other frequencies of the system.
Taking it out of the integrals and setting in (\ref{7}) the lower
frequency bound to $-\infty$, we get delta functions $\delta(\tau
- t \pm x/\upsilon_g)$ from frequency integrals. Thus, from
(\ref{7}) we obtain:
   \begin{equation}\label{9}
\begin{gathered}
  E^+  (x,t) = E_{free}^+
\\
+ \frac{\hbar }{d}\Gamma \theta (x) \sigma_-  \left( t - \frac{x}{v_g} \right)
+ \frac{\hbar }{d}\Gamma \theta ( -x) \sigma_-  \left( t + \frac{x}{v_g} \right),
\end{gathered}
   \end{equation}
where $\Gamma=2\pi g^2(\Omega)$ is the decay rate into the
waveguide modes as obtained by application of Fermi's golden rule
to the Jaynes-Cummings Hamiltonian (\ref{1b}).

In Wigner-Weisskopf approximation the expression for spin operator
$\sigma_-(t)$ reads (see appendix A for detail derivation):
   \begin{equation}\label{10}
\begin{gathered}
  \sigma_- (t) = \sigma_-  (0)e^{ - i(\Omega - i\Gamma )t}
\\
+ ie^{ - i\Omega t} \int\limits_0^\infty  {d\omega } g(\omega
)\int\limits_0^t {d\tau } e^{ - i(\omega  - \Omega )\tau } e^{ -
\Gamma (t - \tau )}
\\  \times \sigma_Z (\tau ) \left({a(\omega ,0)
+ b(\omega ,0) }\right).
\end{gathered}
   \end{equation}
As $[\sigma_Z,H_0]=0$,
$\sigma_Z(\tau)=e^{iH_{JC}\tau}\sigma_Z(0)e^{-iH_{JC}\tau}$.

Replacement of spin operators in (\ref{9}) with the expression
(\ref{10}) yields the final expression for the operator of
electric field:
   \begin{widetext}
   \begin{equation}\label{11}
\begin{gathered}
  E^ +  (x,t) = E_{free}^ +  (x,t)
\\
   + i \frac{\hbar }
{d}\Gamma e^{ - i\Omega (t - \frac{x} {{v_g }})} \theta
(x) \int\limits_0^\infty  {d\omega } g(\omega ) \int\limits_0^{t -
\frac{x} {{v_g }}} {d\tau } e^{ - i(\omega  - \Omega )\tau } e^{ -
\Gamma (t - \frac{x}
{{v_g }} - \tau )} \sigma _Z (\tau ) \left({  a(\omega ,0) + b(\omega ,0) }\right)
\\
   + i \frac{\hbar }
{d} \Gamma e^{ - i\Omega (t + \frac{x} {{v_g }})} \theta ( -
x) \int\limits_0^\infty  {d\omega } g(\omega )\int\limits_0^{t +
\frac{x} {{v_g }}} {d\tau } e^{ - i(\omega  - \Omega )\tau } e^{ -
\Gamma (t + \frac{x}
{{v_g }} - \tau )} \sigma _Z (\tau ) \left({ a(\omega ,0) + b(\omega ,0)}\right)
\\
   + \frac{\hbar }
{d}\Gamma \theta ( - x)e^{ - i(\Omega  - i\Gamma )(t + \frac{x}
{{v_g }})} \sigma _ -  (0) + \frac{\hbar } {d}\Gamma \theta (x)e^{
- i(\Omega  - i\Gamma )(t - \frac{x}
{{v_g }})} \sigma_-  (0).
\end{gathered}
   \end{equation}
   \end{widetext}
The first line in this expression is the electric field in a bare
waveguide (\ref{8}), the next two lines are the electric fields
scattered by the qubit in the forward and backward
directions, respectively, and the last line is the electric field
of spontaneously emitted photon.

In (\ref{9}) we applied the Wigner-Weisskopf approximation in
order to account for the dynamics of the emitter interacting with
the continuum vacuum modes.  This interaction gives rise to an
exponential decay of the qubit from the excited state $|e\rangle$ to the ground
state $|g\rangle$ at a rate $\Gamma$. However, in (\ref{11}) we
keep frequency dependence of the coupling because the application
of this approximation in (\ref{11}) will lead to the divergences
when we treat the two-excitation initial state (Section VI).

The expression (\ref{11}) is the
starting point for the calculation of the matrix elements of the
electric field operator for different initial states of the field.

\section{Photon correlation functions of the qubit-photon system}

\subsection{General expressions for correlation functions}
We define the first and second-order correlation functions by the
conventional expressions \cite{Glauber1963}:
   \begin{equation}\label{g1a}
G^{(1)} (x,t) = \left\langle i \right|E^ -  (x,t)E^ + (x,t)\left|i
\right\rangle,
   \end{equation}
where $\left| i\right\rangle$ is the initial state of a system.
Using a complete set of normalized states  $\left| f\right\rangle
$, the expression (\ref{g1a}) can be rewritten as follows:
   \begin{equation}\label{g1b}
G^{(1)} (x,t) = \sum\limits_f {} \left| {\left\langle f \right|E^
+  (x,t)\left| i \right\rangle } \right|^2.
   \end{equation}
As it follows from (\ref{g1b}), the first order correlation
function is the sum of probability densities of all possible
independent outcomes of the measurement of a field at the moment
$t$ by a detector placed in the point $x$.

The second-order correlation function reads:
   \begin{equation} \label{g2a}
\begin{gathered}
  G^{(2)} (x_2 ,t_2 ;x_1 ,t_1 )
  \\
= \left\langle i \right|E^ -  (x_1 ,t_1 )E^ -  (x_2 ,t_2 )E^ +  (x_2 ,t_2 )E^ +  (x_1 ,t_1 )\left| i \right\rangle .
\end{gathered}
   \end{equation}
With the aid of a complete set of normalized states $\left| f\right\rangle $, the expression (\ref{g2a}) can be rewritten as follows:
   \begin{equation} \label{g2b}
G^{(2)} (x_2 ,t_2 ;x_1 ,t_1 ) = \sum\limits_f {} \left|
{\left\langle f \right|E^ +  (x_2 ,t_2 )E^ +  (x_1 ,t_1 )\left| i
\right\rangle } \right|^2.
   \end{equation}
As it follows from (\ref{g2b}), the second order correlation
function is the sum of probability densities of all possible
independent outcomes resulting from the measurements of a field at
the moments $t_1$  and $t_2$ by two detectors placed in the points
$x_1$ and $x_2$, respectively.

It is worth mentioning here that the initial state $|i\rangle$ is
not necessarily belong to the set $|f\rangle$. For example, it can
be any superposition of the states $|f\rangle$.

Below we denote the full set of the states of the qubit-photon
system as $ \left| {g,n_a ,n_b } \right\rangle ,\left| {e,n_a ,n_b
} \right\rangle $, where the qubit is in the ground or excited
state and there are $n_a$ photons moving right and $n_b$ photons
moving left. The expression for the completeness of the set of
these states is
  \begin{equation}\label{2.0}
\sum\limits_{s,n_a ,n_b } {} \left| {s,n_a ,n_b } \right\rangle
\left\langle {s,n_a ,n_b } \right| = 1,
\end{equation}
where $s=g,e$ and $n_a,n_b=0,1,2...$.

The vacuum state is $ \left| {g,0_a ,0_b } \right\rangle$: $E^ +
(x,t)\left| {g,0_a ,0_b } \right\rangle  = 0 $. Every state has a
definite number of excitations. There are $n_a+n_b$ excitations
for the state $ \left| {g,n_a ,n_b } \right\rangle$ and
$n_a+n_b+1$ excitations for the state $ \left| {e,n_a ,n_b }
\right\rangle$. The action of the operator $E^+(x,t)$ on these
states removes one excitation from the system. For example, the
action $ E^+(x,t)\left| {e,n_a ,n_b } \right\rangle$ leads to
three possible output states $\left| {g,n_a ,n_b } \right\rangle
,\;\left| {e,n_a  - 1,n_b } \right\rangle ,\;\left| {e,n_a, n_b -1
} \right\rangle $. Therefore, there will be only three non-zero
matrix elements: $\left\langle {g,n_a ,n_b } \right|E^ +
(x,t)\left| {e,n_a ,n_b } \right\rangle$, $\left\langle {e,n_a  -
1,n_b } \right|E^ + (x,t)\left| {e,n_a ,n_b } \right\rangle$,
$\left\langle {e,n_a ,n_b - 1} \right|E^ + (x,t)\left| {e,n_a ,n_b
} \right\rangle$. Each of these matrix elements is the probability
amplitude for the corresponding transition and can be considered
as a photon wavefunction associated with the specific transitions.
They describe the detection in the point $x$ at the moment $t$ of
the spontaneously emitted photon, right moving photon, and left
moving photon, respectively. In this case, the measurement
outcomes can be considered as independent events, therefore the
full probability density is given by a sum of squared moduli of
these matrix elements:
   \begin{equation}\label{2.1}
\begin{gathered}
p(x,t) = G^{(1)}(x,t) = \left| {\left\langle {g,n_a ,n_b } \right|E^+  (x,t)\left| {e,n_a ,n_b } \right\rangle } \right|^2
\\
   + \left| {\left\langle {e,n_a  - 1,n_b } \right|E^+  (x,t)\left| {e,n_a ,n_b } \right\rangle } \right|^2
\\
+ \left| {\left\langle {e,n_a ,n_b  - 1} \right|E^+  (x,t)\left| {e,n_a ,n_b } \right\rangle } \right|^2.
\end{gathered}
   \end{equation}
The equation (\ref{2.1}) can be rewritten in the conventional form
of the first-order correlation function:
   \begin{equation}\label{2.2}
G^{(1)} (x,t) =  {\left\langle {e,n_a ,n_b } \right|E^ - (x,t)E^ +
(x,t)\left| {e,n_a ,n_b } \right\rangle }.
   \end{equation}
The equation (\ref{2.1}) follows immediately from (\ref{2.2}) if we use the completeness condition (\ref{2.0}) between operators $E^-(x,t)$ and $E^+(x,t)$ in (\ref{2.2}) and omit all elements that are zero.

The correlation function $G^{(1)}(x,t)$ describes a process in which the photon field produced by the qubit-waveguide system in the initial state $\left|e,n_a,n_b\right > $ is detected at the point $x$ and at the time moment $t$.

For the initial state $|i\rangle=|e,n_a,n_b\rangle$ the
correlation function of the second order is defined as follows
   \begin{widetext}
  \begin{equation}\label{2.3}
G^{(2)} (x_2 ,t_2 ;x_1 ,t_1 ) = \left\langle {e,n_a ,n_b }
\right|E^ -  (x_1 ,t_1 )E^ -  (x_2 ,t_2 )E^ +  (x_2 ,t_2 )E^ +
(x_1 ,t_1 )\left| {e,n_a ,n_b } \right\rangle.
  \end{equation}
   \end{widetext}
This correlation function describes the process of the
measurements with two photon counters located at the points $x_1$
and $x_2$. A first counter detects  photon at time $t_1$ and a second
counter detects photon at $t_2$. It is worth noting that in
general the fields $E^+(x_1,t_1)$ and $E^+(x_2,t_2)$ in
(\ref{2.3}) do not commute (see expression (\ref{11})). Therefore,
the outcomes of the measurements of two detectors should
inevitably be correlated.

\subsection{Single-photon subspace}

In what follows we assume that there are no left moving photons in
the initial state, $n_b=0$, and there is a single right moving
photon, $n_a=1$. Therefore, we consider here two kinds of the states: two states with a single excitation, $  \left| {g,1_a ,0_b
} \right\rangle $ and $  \left| {e,0_a ,0_b } \right\rangle $, and
the state with two excitations $ \left| {e,1_a ,0_b }
\right\rangle $. As the action of $b(\omega,0)$ on these states is
zero, we may omit the terms with $b(\omega,0)$ from (\ref{11}).
For  convenience we denote these states as $ \left| {e,0 }
\right\rangle $, $ \left| {g,1 } \right\rangle $ and $\left| {e,1}
\right\rangle $, assuming implicitly that initially there is no
left moving photon and omitting subscript $a$ for right moving
photon.

For multimode case the state $  \left| {s,1 } \right\rangle $,
where $s=g,e$ consists of infinite number of states each of which
has a single photon in $k$th mode:
\begin{equation}\label{s}
    \{ \left| {s,1} \right\rangle \}  = \left\{ \begin{gathered}
  \left| {s,1_{k_1 } } \right\rangle  \equiv \left| {s,1_{k_1 } ,0,0,0......0} \right\rangle , \hfill \\
  \left| {s,1_{k_2 } } \right\rangle  \equiv \left| {s,0,1_{k_2 } ,0,0,0.....0} \right\rangle , \hfill \\
  \left| {s,1_{k3} } \right\rangle  \equiv \left| {s,0,0,1_{k_3 } ,0,0,0......0} \right\rangle , \hfill \\
  ................................... \hfill \\
\end{gathered}  \right.
   \end{equation}
The completeness relation can be written as follows:
   \begin{equation}\label{comp}
\begin{gathered}
  \left| {g,0} \right\rangle \left\langle {g,0} \right| + \left| {e,0} \right\rangle \left\langle {e,0} \right| + \left| {g,1} \right\rangle \left\langle {g,1} \right| + \left| {e,1} \right\rangle \left\langle {e,1} \right|
\\
   + \sum\limits_{s = g,e} {} \sum\limits_{n = 2}^\infty  {} \left| {s,n} \right\rangle \left\langle {s,n} \right| = 1,
\end{gathered}
   \end{equation}
where we explicitly write the single photon terms.

For multimode continuum the quantity $|g,1\rangle\langle g,1|$
reads
\begin{equation}\label{g}
\left| {g,1} \right\rangle \left\langle {g,1} \right| =
\int\limits_0^\infty  {d\omega } a^\dagger  (\omega ,0)\left| {g,0}
\right\rangle \left\langle {g,0} \right|a(\omega ,0).
\end{equation}

\subsection{Single-excitation initial state}

Here we consider the initial states $|i\rangle$ with one
excitation, the state $  \left| {e,0 } \right\rangle $ and the
state $ \left| {G,1 } \right\rangle $, which is a linear
superposition of basis single-photon states (\ref{s}):
   \begin{equation}\label{4.2a}
\left| {G,1} \right\rangle  = \int\limits_0^\infty  {d\omega }
f(\omega )a^\dagger  (\omega ,0)\left| {g,0} \right\rangle,
   \end{equation}
where $f(\omega)$ is a frequency spectrum of incident photon pulse, which satisfies the normalizing condition $\int\limits_0^\infty  {d\omega } \left| {f(\omega )} \right|^2  =1 $.

The state $|G,1\rangle$ describes the initial photon pulse incident from the left with a qubit being in the ground state. Note that this pulse can have any arbitrary shape since the function $f(\omega)$ is not specified. Later on, we use plane wave in a form of a delta-pulse to simplify the results.

The action of $E^+(x,t)$ on these states leads to the only final
state $ \left| {g,0 } \right\rangle $. Therefore, there is one
nonzero matrix element for each of initial states: $\left\langle
{g,0 } \right|E^ + (x,t)\left| {G,1 } \right\rangle$ and
$\left\langle {g,0 } \right|E^ + (x,t)\left| {e,0 }
\right\rangle$. In this case, the first order correlation
functions are as follows
   \begin{equation}\label{2.4}
    G^{(1)}(x,t)=\left|\left\langle {g,0 } \right|E^
+ (x,t)\left| {G,1 } \right\rangle\right|^2,
   \end{equation}
and
   \begin{equation}\label{2.4a}
    G^{(1)}(x,t)=\left|\left\langle {g,0 } \right|E^
+ (x,t)\left| {e,0 } \right\rangle\right|^2.
   \end{equation}

As $\left|g,0\right\rangle$ is the vacuum state, in this case, the
second-order correlation function is zero.

We show below, that the first-order correlation function
(\ref{2.4}) describes the probability densities of transmitted and
reflected spectra from unexcited qubit, while the expression
(\ref{2.4a}) describes the probability density of spontaneously
emitted photon.

\subsection{Two-excitation initial state}

Here we consider the initial state $ \left| E,1 \right\rangle $
having two excitations:
   \begin{equation}\label{4.8a}
\left| {E,1} \right\rangle  = \int\limits_0^\infty  {d\omega }
f(\omega )a^\dagger  (\omega ,0)\left| {e,0} \right\rangle.
   \end{equation}
The state $|E,1\rangle$ describes the initial photon incident from
the left with a qubit being in the excited state.

The action of $E^+(x_,t)$ on this state leads to two final states
with one excitation, $ \left| {g,1 } \right\rangle $ and  $ \left|
{e,0 } \right\rangle $. Therefore, in this case, the correlation
function of the first order reads:
   \begin{equation}\label{2.5}
\begin{gathered}
  G^{(1)} (x,t) = \left| {\left\langle {e,0} \right|E^ +  (x,t)\left| {E,1} \right\rangle } \right|^2
\\
   + \left\langle {E,1} \right|E^ -  (x,t)\left| {g,1} \right\rangle \left\langle {g,1} \right|E^ +  (x,t)\left| {E,1} \right\rangle.
\end{gathered}
   \end{equation}
Using multimode continuum expression (\ref{g}), the equation
(\ref{2.5}) takes the form:
   \begin{equation}\label{2.5a}
    \begin{gathered}
  G^{(1)} (x,t) = \left| {\left\langle {e,0} \right|E^ +  (x,t)\left| {E,1} \right\rangle } \right|^2
\\
   + \int\limits_0^\infty  {d\omega } \left| {\left\langle {g,0} \right|a(\omega ,0)E^ +  (x,t)\left| {E,1} \right\rangle } \right|^2.
\end{gathered}
   \end{equation}

The consecutive application of two field operators to the initial
state $|E,1\rangle$,
$E^+(x_2,t_2)E^+(x_1,t_1)\left|E,1\right\rangle$ removes two
excitations from the system, leaving the only final state
$\left|g,0\right\rangle$. In this case, the expression for the
second-order correlation function is as follows:
   \begin{equation}\label{2.6}
G^{(2)} (x_2 ,t_2 ;x_1 ,t_1 ) = \left| {\left\langle {g,0}
\right|E^ +  (x_2 ,t_2)E^ +  (x_1 ,t_1)\left| {E,1} \right\rangle
} \right|^2.
   \end{equation}

Applying the completeness relation (\ref{2.0}) between two field
operators in (\ref{2.6}) and leaving only non-zero matrix
elements, we obtain for second-order correlation function:

\begin{widetext}
   \begin{equation}\label{2.7}
\begin{gathered}
  G^{(2)} (x_1 ,t_1 ;x_2 ,t_2 ) = \left| {\left\langle {g,0} \right|E^ +  (x_2 ,t_2 )\left| {g,1} \right\rangle \left\langle {g,1} \right|E^ +  (x_1 ,t_1 )\left| {E,1} \right\rangle  + \left\langle {g,0} \right|E^ +  (x_2 ,t_2 )\left| {e,0} \right\rangle \left\langle {e,0} \right|E^ +  (x_1 ,t_1 )\left| {E,1} \right\rangle } \right|^2
\\
   = \left| {\int\limits_0^\infty  {d\omega } \left\langle {g,0} \right|E^ +  (x_2 ,t_2 )a^\dagger  (\omega ,0)\left| {g,0} \right\rangle \left\langle {g,0} \right|a(\omega ,0)E^ +  (x_1 ,t_1 )\left| {E,1} \right\rangle  + \left\langle {g,0} \right|E^ +  (x_2 ,t_2 )\left| {e,0} \right\rangle \left\langle {e,0} \right|E^ +  (x_1 ,t_1 )\left| {E,1} \right\rangle } \right|^2.
\end{gathered}
   \end{equation}
The second-order correlation function (\ref{2.7}) can be written
as follows:
   \begin{equation}\label{2.8}
G^{(2)} (x_1 ,t_1 ;x_2 ,t_2 ) = G_{path1}^{(2)}  + G_{path2}^{(2)}
+ G^{(2)}_{int},
   \end{equation}
where:
   \begin{subequations}\label{2.9}
\begin{align}
 G_{path1}^{(2)}  = &\left| {\int\limits_0^\infty  {d\omega }
\left\langle {g,0} \right|E^ +  (x_2 ,t_2 )a^\dagger  (\omega ,0)\left|
{g,0} \right\rangle \left\langle {g,0} \right|a(\omega ,0)E^+
(x_1 ,t_1 )\left| {E,1} \right\rangle } \right|^2,  \label{2.9a}
\\
&G_{path2}^{(2)}  = \left| {\left\langle {g,0} \right|E^ +  (x_2 ,t_2 )\left| {e,0} \right\rangle } \right|^2 \left| {\left\langle {e,0} \right|E^ +  (x_1 ,t_1 )\left| {E,1} \right\rangle } \right|^2,  \label{2.9b}
\end{align}
   \end{subequations}
   \begin{equation}\label{2.10}
    \begin{gathered}
\begin{gathered}
  G_{\operatorname{int} }^{(2)}  = \int\limits_0^\infty  {d\omega } \left\langle {g,0} \right|E^ +  (x_2 ,t_2 )a^\dagger  (\omega ,0)\left| {g,0} \right\rangle \left\langle {g,0} \right|a(\omega ,0)E^ +  (x_1 ,t_1 )\left| {E,1} \right\rangle \left\langle {e,0} \right|E^ -  (x_2 ,t_2 )\left| {g,0} \right\rangle \left\langle {E,1} \right|E^ -  (x_1 ,t_1 )\left| {e,0} \right\rangle
\\
   + \left\langle {g,0} \right|E^ +  (x_2 ,t_2 )\left| {e,0} \right\rangle \left\langle {e,0} \right|E^ +  (x_1 ,t_1 )\left| {E,1} \right\rangle \int\limits_0^\infty  {d\omega} \left\langle {g,0} \right| a(\omega ,0) E^-  (x_2 ,t_2) \left| {g,0} \right\rangle \left\langle {E,1} \right|E^- (x_1 ,t_1 ) a^\dagger (\omega ,0)\left| {g,0} \right\rangle.
\end{gathered}
\end{gathered}
   \end{equation}
\end{widetext}

The interference term  (\ref{2.10}) describes the interference
between two photons which comes to detectors $D_1$ and $D_2$ by two different pathways (see
Fig.1). The detector $D_1$ clicks first
as it is located closer to the qubit, $|x_1| < |x_2|$. The
evolution of the initial state $|E,1\rangle$ gives rise to two
photons, the spontaneously emitted photon and the photon of the
scattered field. One possible pathway of the measurements by two
detectors is when the spontaneous photon first reaches the
detector $D_1$ leaving the system in the state $|g,1\rangle$, and
later the scattered photon reaches the detector $D_2$ leaving the
system in the state $|g,0\rangle$. The photon probability
amplitude which describes this process is $\left\langle {g,0}
\right|E^ +  (x_2 ,t_2 )\left| {g,1} \right\rangle
  \left\langle {g,1} \right|E^ +  (x_1 ,t_1 )\left| {E,1} \right\rangle$.
Alternative possible pathway is when the scattered photon first
reaches the detector $D_1$ leaving the system in the state
$|e,0\rangle$, and later the spontaneous photon reaches the
detector $D_2$ leaving the system in the state $|g,0\rangle$. The
photon probability amplitude which describes this process is $\left\langle {g,0} \right|E^ +  (x_2 ,t_2 )\left|
{e,0}\right\rangle \left\langle {e,0} \right|E^ +  (x_1 ,t_1
)\left| {E,1} \right\rangle $. As these two pathways are
undistinguishable, the two probability amplitudes must first be
summed up and then squared modulus of this sum has to be
calculated.

   \begin{figure}
\includegraphics[width=8 cm]{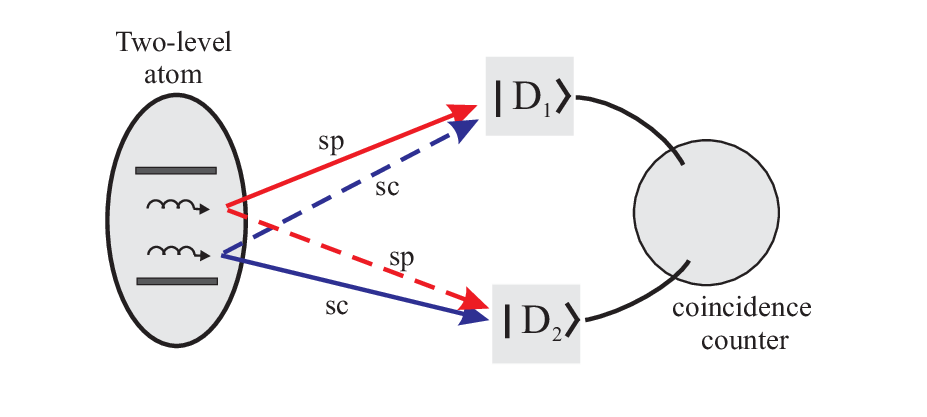}\\
\caption{A schematic diagram for a two-photon interference.
 First pathway (solid arrows): the spontaneously emitted photon $\textit{sp}$
is measured by detector $D_1$ and then the scattering photon $\textit{sc}$
is measured by detector $D_2$. The other pathway (dashed arrows): the scattering
photon $\textit{sc}$ first is measured by detector $D_1$ and then the
spontaneous emited photon $\textit{sp}$ is measured by detector $D_2$.}
  \label{fig0}
  \end{figure}

\section{Calculation of the matrix elements of the electric field operator for the single excitation initial states}

\subsection{Initial state $  \left| {e,0 } \right\rangle $}

From expression (\ref{11}) it is easily seen that the contribution
to the matrix element $\left\langle {g,0 } \right|E^ + (x,t)\left|
{e,0 } \right\rangle$ comes only from the last line in (\ref{11})
which describes the spontaneous decay of the excited atom:
   \begin{equation}\label{4.1}
\begin{gathered}
\left\langle {g,0} \right|E^ +  (x,t)\left| {e,0} \right\rangle  =
\\
\frac{\hbar }
{d}\Gamma \theta ( - x)e^{ - i(\Omega  - i\Gamma )(t + \frac{x}
{v_g })}  + \frac{\hbar } {d}\Gamma \theta (x)e^{ - i(\Omega  -
i\Gamma )(t - \frac{x}
{v_g })}.
\end{gathered}
   \end{equation}
This expression contributes directly to the first-order
correlation function (\ref{2.4a}) and describes a travelling of
the spontaneously emitted photon in two opposite directions.

As is seen from (\ref{2.7}), the matrix element (\ref{4.1})
contributes also to the second-order correlation function for the
initial state $\left|E,1\right\rangle$.

\subsection{Initial state $  \left| {G,1 } \right\rangle $}

In this subsection we calculate the matrix element $\left\langle
{g,0 } \right|E^ + (x,t)\left| {G,1 } \right\rangle$ which
contributes directly to the first-order correlation function
(\ref{2.4}) and show that it leads to the correct expressions for
the transmitted and reflected photon fields.

The details of the calculations are given in the Appendix B. Here
we write down the final result:
   \begin{equation}\label{4.3}
\begin{gathered}
  \left\langle {g,0} \right|E^+  (x,t)\left| {G,1} \right\rangle  = i\frac{\hbar }
{d}\int\limits_0^\infty  {d\omega } g(\omega )f(\omega )e^{ -
i\omega \left( {t - \frac{x}
{v_g }} \right)}
\\
+ \frac{\hbar }
{d}\Gamma \theta (x)\int\limits_0^\infty  {d\omega } g(\omega
)f(\omega )\frac{{e^{ - i\omega \left( {t - \frac{x} {v_g }}
\right)}  - e^{ - i(\Omega  - i\Gamma )\left( {t - \frac{x} {v_g
}} \right)} }}
{{\omega  - \Omega  + i\Gamma }}
\\
   + \frac{\hbar }
{d}\Gamma \theta ( - x)\int\limits_0^\infty  {d\omega } g(\omega
)f(\omega )\frac{{e^{ - i\omega \left( {t + \frac{x} {v_g }}
\right)}  - e^{ - i(\Omega  - i\Gamma ) \left( {t + \frac{x} {v_g
}} \right)} }}
{{\omega  - \Omega  + i\Gamma }}.
\end{gathered}
   \end{equation}
The first term in the right-hand side in (\ref{4.3}) describes the
initial photon incident from the left, the second term describes
the transmitted photon, and the third term describes the reflected
photon.

For $t\rightarrow\infty$ we obtain from (\ref{4.3}) transmitted signal:
   \begin{equation}\label{4.4}
\begin{gathered}
  \left\langle {g,0} \right|E^ +  (x,t)\left| {G,1} \right\rangle _{x > 0,t \to \infty } =
 \\
   = i\frac{\hbar }
{d}\int\limits_0^\infty  {d\omega } g(\omega )f(\omega
)\frac{{\omega  - \Omega }} {{\omega  - \Omega  + i\Gamma }}e^{ -
i\omega \left( {t - \frac{x}
{{v_g }}} \right)},
\end{gathered}
   \end{equation}
and reflected signal:
   \begin{equation}\label{4.5}
\begin{gathered}
  \left\langle {g,0} \right|E^ +  (x,t)\left| {G,1} \right\rangle _{x < 0,t \to \infty } =
\\
   = \frac{\hbar }
{d}\int\limits_0^\infty  {d\omega } g(\omega )f(\omega
)\frac{\Gamma } {{\omega  - \Omega  + i\Gamma }}e^{ - i\omega
\left( {t + \frac{x}
{{v_g }}} \right)}.
\end{gathered}
   \end{equation}

These expressions are valid for any photon waveform $f(\omega)$.
In principle we may consider the incident photon as a plane wave
with a fixed frequency $\omega_0$, which can be approximated by a
narrow  Gaussian packet \cite{Green2023a}. In this case, we may
approximate $f(\omega)$ as
$f(\omega)=\sqrt{\Delta}\delta(\omega-\omega_0)$, where $\Delta $
is the width of a narrow Gaussian packet in the frequency domain.
For this plane wave approximation we obtain from (\ref{4.4}) and
(\ref{4.5}) the expressions for transmitted and reflected photon
fields:
  \begin{subequations}
   \begin{align} \label{4.6}
\begin{gathered}
  \left\langle {g,0} \right|E^ +  (x,t)\left| {G,1} \right\rangle _{x > 0 ,t \to \infty }
\\
   = i\frac{\hbar }
{d}g(\omega _0 )\sqrt \Delta  \frac{{\omega _0  - \Omega }}
{{\omega _0  - \Omega  + i\Gamma }}e^{ - i\omega _0 \left( {t -
\frac{x} {v_g }} \right)},
\end{gathered}
   \end{align}
   \begin{align} \label{4.7}
\begin{gathered}
  \left\langle {g,0} \right|E^ +  (x,t)\left| {G,1} \right\rangle
   _{x < 0,t \to \infty }
\\
   = \frac{\hbar }
{d}g(\omega _0 )\sqrt \Delta  \frac{\Gamma } {{\omega _0  - \Omega
+ i\Gamma }}e^{ - i\omega _0 \left( {t + \frac{x} {v_g }}
\right)}.
\end{gathered}
   \end{align}
   \end{subequations}
The frequency dependence of transmitted and reflected fields in
(\ref{4.6}) and (\ref{4.7}) coincides with well-known results for
transmitted and reflected spectra obtained in the frame of
stationary, time-independent scattering theories \cite{Shen05a,
Shen05b}. The squared moduli of (\ref{4.6}) and (\ref{4.7}), which
are exactly the first-order correlation function (\ref{2.4}),
provide the measured probability densities for reflected and
transmitted fields.

\section{Calculation of the matrix elements of the electric field operator for the two-excitation initial state}

\subsection{First-order correlation function $G^{(1)}(x,t)$}

First we calculate the action of $E^+(x,t)$ on the initial state
$\left|E,1\right\rangle$. The result is as follows (details of the
calculations are given in Appendix C):
   \begin{widetext}
   \begin{equation}\label{4.9}
\begin{gathered}
  E^ +  (x,t)\left| {E,1} \right\rangle  = i\frac{\hbar }
{d}\int\limits_0^\infty  {d\omega } g(\omega )f(\omega )e^{ -
i\omega \left( {t - \frac{x}
{{v_g }}} \right)} \left| {e,0} \right\rangle
\\
   + i\frac{\hbar }
{d}\Gamma e^{ - i(\Omega  - i\Gamma )(t - \frac{x} {{v_g }})}
\theta (x)\int\limits_0^\infty  {d\omega } g(\omega )f(\omega
)\int\limits_0^{t - \frac{x} {{v_g }}} {d\tau } e^{ - i(\omega  -
\Omega  + i\Gamma )\tau } \left( {\cos \left( {2\sqrt \Lambda
\tau } \right)\left| {e,0} \right\rangle  + \frac{i}
{{2\sqrt \Lambda  }}\sin \left( {2\sqrt \Lambda  \tau } \right)\overline {\left| {G,1} \right\rangle } } \right)
\\
   + i\frac{\hbar }
{d}\Gamma e^{ - i(\Omega  - i\Gamma )(t + \frac{x} {{v_g }})}
\theta ( - x)\int\limits_0^\infty  {d\omega } g(\omega )f(\omega
)\int\limits_0^{t + \frac{x} {{v_g }}} {d\tau } e^{ - i(\omega  -
\Omega  + i\Gamma )\tau } \left( {\cos \left( {2\sqrt \Lambda
\tau } \right)\left| {e,0} \right\rangle  + \frac{i}
{{2\sqrt \Lambda  }}\sin \left( {2\sqrt \Lambda  \tau } \right)\overline {\left| {G,1} \right\rangle } } \right)
\\
   + \frac{\hbar }
{d}\Gamma \theta ( - x)e^{ - i(\Omega  - i\Gamma )(t + \frac{x}
{{v_g }})} \left| {G,1} \right\rangle  + \frac{\hbar } {d}\Gamma
\theta (x)e^{ - i(\Omega  - i\Gamma )(t - \frac{x}
{{v_g }})} \left| {G,1} \right\rangle,
\end{gathered}
   \end{equation}
where
   \begin{equation}\label{4.10}
\Lambda  =  {\int\limits_0^\infty  {d\omega } g^2 (\omega )} ,
   \end{equation}
   \begin{equation}\label{4.11}
\overline {\left| {G,1} \right\rangle }  = \int\limits_0^\infty
{d\omega } g(\omega )a^\dagger (\omega ,0)\left| {g,0}
\right\rangle.
   \end{equation}
The state (\ref{4.11}) can be interpret the same way as state $\left| G, 1 \right\rangle $ in (\ref{4.2a}) – it is a linear superposition of basic single-photon states, but this time photons are corresponds to spontaneous emission due to qubit decay, and not to incident fields.

As is seen from (\ref{4.9}), the action of $E^+(x,t)$ on the state
$\left|E,1\right\rangle$ leaves the field in a superposition of
state $\left|e,0\right\rangle$ and single photon Fock states
(\ref{s}).

From (\ref{4.9}) we obtain the matrix elements which contribute to
the first-order correlation function (\ref{2.5a}):
   \begin{equation}\label{4.12}
\begin{gathered}
  \left\langle {e,0} \right|E^ +  (x,t)\left| {E,1} \right\rangle  = i\frac{\hbar }
{d}\int\limits_0^\infty  {d\omega } g(\omega )f(\omega )e^{ -
i\omega \left( {t - \frac{x}
{{v_g }}} \right)}
\\
   + i\frac{\hbar }
{d}\Gamma e^{ - i(\Omega  - i\Gamma )(t - \frac{x} {{v_g }})}
\theta (x)\int\limits_0^\infty  {d\omega } g(\omega )f(\omega
)\int\limits_0^{t - \frac{x}
{{v_g }}} {d\tau } e^{ - i(\omega  - \Omega  + i\Gamma )\tau } \cos \left( {2\sqrt \Lambda  \tau } \right)
\\
   + i\frac{\hbar }
{d}\Gamma e^{ - i(\Omega  - i\Gamma )(t + \frac{x} {{v_g }})}
\theta ( - x)\int\limits_0^\infty  {d\omega } g(\omega )f(\omega
)\int\limits_0^{t + \frac{x}
{{v_g }}} {d\tau } e^{ - i(\omega  - \Omega  + i\Gamma )\tau } \cos \left( {2\sqrt \Lambda  \tau } \right),
\end{gathered}
   \end{equation}

   \begin{equation} \label{4.13}
\begin{gathered}
  \left\langle {g,0} \right|a(\omega ',0)E^ +  (x,t)\left| {E,1} \right\rangle  =
\\
   - \frac{\hbar }
{d}\Gamma e^{ - i(\Omega  - i\Gamma )(t - \frac{x} {{v_g }})}
\theta (x)\int\limits_0^\infty  {d\omega } g(\omega )f(\omega
)\int\limits_0^{t - \frac{x} {{v_g }}} {d\tau } e^{ - i(\omega  -
\Omega  + i\Gamma )\tau } \frac{1}
{{2\sqrt \Lambda  }}\sin \left( {2\sqrt \Lambda  \tau } \right)g(\omega ')
\\
   - \frac{\hbar }
{d}\Gamma e^{ - i(\Omega  - i\Gamma )(t + \frac{x} {{v_g }})}
\theta ( - x)\int\limits_0^\infty  {d\omega } g(\omega )f(\omega
)\int\limits_0^{t + \frac{x} {{v_g }}} {d\tau } e^{ - i(\omega  -
\Omega  + i\Gamma )\tau } \frac{1}
{{2\sqrt \Lambda  }}\sin \left( {2\sqrt \Lambda  \tau } \right)g(\omega ')
\\
   + \frac{\hbar }
{d}\Gamma \theta ( - x)e^{ - i(\Omega  - i\Gamma )(t + \frac{x}
{{v_g }})} f(\omega ') + \frac{\hbar } {d}\Gamma \theta (x)e^{ -
i(\Omega  - i\Gamma )(t - \frac{x}
{{v_g }})} f(\omega ').
\end{gathered}
   \end{equation}

The calculation of the integrals over $\tau$ in expressions
(\ref{4.12}) and (\ref{4.13}) yields the following result:
   \begin{equation}\label{4.15}
\begin{gathered}
  \left\langle {e,0} \right|E^ +  (x,t)\left| {E,1} \right\rangle  = i\frac{\hbar }
{d}\int\limits_0^\infty  {d\omega } g(\omega )f(\omega )e^{ -
i\omega \left( {t - \frac{x}
{v_g }} \right)}
\\
   - \frac{\hbar }
{d}\frac{\Gamma } {2}\int\limits_0^\infty  {d\omega }
\frac{{g(\omega )f(\omega )}} {{\omega  - \Omega  - 2\sqrt \Lambda
+ i\Gamma }}\left( {e^{ - i(\omega  - 2\sqrt \Lambda  )\left( {t -
\frac{x} {v_g }} \right)}  - e^{ - i(\Omega  - i\Gamma )\left(
{t - \frac{x} {{v_g }}} \right)} } \right)\theta (x)\theta \left(
{t - \frac{x}
{v_g }} \right)
\\
   - \frac{\hbar }
{d}\frac{\Gamma } {2}\int\limits_0^\infty  {d\omega }
\frac{{g(\omega )f(\omega )}} {{\omega  - \Omega  + 2\sqrt \Lambda
+ i\Gamma }}\left( {e^{ - i(\omega  + 2\sqrt \Lambda  )\left( {t -
\frac{x} {{v_g }}} \right)}  - e^{ - i(\Omega  - i\Gamma )\left(
{t - \frac{x} {{v_g }}} \right)} } \right)\theta (x)\theta \left(
{t - \frac{x}
{v_g }} \right)
\\
   - \frac{\hbar }
{d}\frac{\Gamma } {2}\int\limits_0^\infty  {d\omega }
\frac{{g(\omega )f(\omega )}} {{\omega  - \Omega  + 2\sqrt \Lambda
+ i\Gamma }}\left( {e^{ - i(\omega  + 2\sqrt \Lambda  )\left( {t +
\frac{x} {{v_g }}} \right)}  - e^{ - i(\Omega  - i\Gamma )\left(
{t + \frac{x} {{v_g }}} \right)} } \right)\theta ( - x)\theta
\left( {t + \frac{x}
{{v_g }}} \right)
\\
   - \frac{\hbar }
{d}\frac{\Gamma } {2}\int\limits_0^\infty  {d\omega }
\frac{{g(\omega )f(\omega )}} {{\omega  - \Omega  - 2\sqrt \Lambda
+ i\Gamma }}\left( {e^{ - i(\omega  - 2\sqrt \Lambda  )\left( {t +
\frac{x} {{v_g }}} \right)}  - e^{ - i(\Omega  - i\Gamma )\left(
{t + \frac{x} {{v_g }}} \right)} } \right)\theta ( - x)\theta
\left( {t + \frac{x}
{{v_g }}} \right),
\end{gathered}
   \end{equation}

   \begin{equation} \label{4.16}
\begin{gathered}
  \left\langle {g,0} \right|a({\omega',0})E^ +  (x,t)\left| {E,1} \right\rangle
\\
   =  - \frac{\hbar }
{d}\frac{\Gamma }
{4}\frac{g(\omega')}{\sqrt{\Lambda}}\int\limits_0^\infty {d\omega
} \frac{{g(\omega )f(\omega )}} {{\omega  - \Omega  - 2\sqrt
\Lambda + i\Gamma }}\left( {e^{ - i(\omega  - 2\sqrt \Lambda
)\left( {t - \frac{x} {{v_g }}} \right)}  - e^{ - i(\Omega -
i\Gamma )\left( {t - \frac{x} {v_g }} \right)} } \right)\theta
(x)\theta \left( {t - \frac{x} {v_g }} \right)
\\
   + \frac{\hbar }
{d}\frac{\Gamma }
{4}\frac{g(\omega')}{\sqrt{\Lambda}}\int\limits_0^\infty {d\omega
} \frac{{g(\omega )f(\omega )}} {{\omega  - \Omega  + 2\sqrt
\Lambda + i\Gamma }}\left( {e^{ - i(\omega  + 2\sqrt \Lambda
)\left( {t - \frac{x} {v_g }} \right)}  - e^{ - i(\Omega - i\Gamma
)\left( {t - \frac{x} {v_g }} \right)} } \right)\theta (x)\theta
\left( {t - \frac{x} {{v_g }}} \right)
\\
   + \frac{\hbar }
{d}\frac{\Gamma }
{4}\frac{g(\omega')}{\sqrt{\Lambda}}\int\limits_0^\infty {d\omega
} \frac{{g(\omega )f(\omega )}} {{\omega  - \Omega  + 2\sqrt
\Lambda + i\Gamma }}\left( {e^{ - i(\omega  + 2\sqrt \Lambda
)\left( {t + \frac{x} {v_g }} \right)}  - e^{ - i(\Omega - i\Gamma
)\left( {t + \frac{x} {v_g }} \right)} } \right)\theta ( -
x)\theta \left( {t + \frac{x} {{v_g }}} \right)
\\
   - \frac{\hbar }
{d}\frac{\Gamma }
{4}\frac{g(\omega')}{\sqrt{\Lambda}}\int\limits_0^\infty {d\omega
} \frac{{g(\omega )f(\omega )}} {{\omega  - \Omega  - 2\sqrt
\Lambda + i\Gamma }}\left( {e^{ - i(\omega  - 2\sqrt \Lambda
)\left( {t + \frac{x} {{v_g }}} \right)}  - e^{ - i(\Omega  -
i\Gamma )\left( {t + \frac{x} {{v_g }}} \right)} } \right)\theta (
- x)\theta \left( {t + \frac{x} {{v_g }}} \right)
\\
   + \frac{\hbar }
{d}\Gamma f(\omega')\theta ( - x)e^{ - i(\Omega  - i\Gamma )(t +
\frac{x} {{v_g }})}  + \frac{\hbar } {d}\Gamma f(\omega') \theta
(x)e^{ - i(\Omega  - i\Gamma )(t - \frac{x}{v_g })},
\end{gathered}
   \end{equation}

From the matrix elements (\ref{4.15}), (\ref{4.16}) we can
construct the first-order correlation function (\ref{2.5a}). In
the plane wave approximation
$f(\omega)=\sqrt{\Delta}\delta(\omega-\omega_0)$, we obtain from
(\ref{2.5a}) the transmission and reflection spectra in the limit
$t\rightarrow\infty$:
   \begin{equation}\label{4.25}
\begin{gathered}
  G^{(1)} (x,t)_{x > 0,t \to \infty }  = \int\limits_0^\infty  {d\omega '} \left| {\left\langle {g,0} \right|a(\omega ',0)E^ +  (x,t)\left| {E,1} \right\rangle _{x > 0,t \to \infty } } \right|^2  + \left| {\left\langle {e,0} \right|E^ +  (x,t)\left| {E,1} \right\rangle _{x > 0,t \to \infty } } \right|^2
\\
   = \frac{{\hbar ^2 }}
{{d^2 }}g^2 (\omega _0 )\Delta \left| {\left( {1 + i\frac{\Gamma }
{2}\frac{{e^{i2\sqrt \Lambda  )\left( {t - \frac{x} {{v_g }}}
\right)} }} {{\omega _0  - \Omega  - 2\sqrt \Lambda   + i\Gamma }}
+ i\frac{\Gamma } {2}\frac{{e^{ - i2\sqrt \Lambda  \left( {t -
\frac{x} {{v_g }}} \right)} }}
{{\omega _0  - \Omega  + 2\sqrt \Lambda   + i\Gamma }}} \right)} \right|^2
\\
   + \frac{{\hbar ^2 }}
{{d^2 }}\frac{{\Gamma ^2 }} {{16}}g^2 (\omega _0 )\Delta \left|
{\frac{{e^{i2\sqrt \Lambda  \left( {t - \frac{x} {{v_g }}}
\right)} }} {{\omega _0  - \Omega  - 2\sqrt \Lambda   + i\Gamma }}
- \frac{{ - i2\sqrt \Lambda  \left( {t - \frac{x} {{v_g }}}
\right)}}
{{\omega _0  - \Omega  + 2\sqrt \Lambda   + i\Gamma }}} \right|^2,
\end{gathered}
   \end{equation}
   \begin{equation}\label{4.26}
\begin{gathered}
  G^{(1)} (x,t)_{x < 0,t \to \infty }  = \int\limits_0^\infty  {d\omega '} \left| {\left\langle {g,0} \right|a(\omega ',0)E^ +  (x,t)\left| {E,1} \right\rangle _{x < 0,t \to \infty } } \right|^2  + \left| {\left\langle {e,0} \right|E^ +  (x,t)\left| {E,1} \right\rangle _{x < 0,t \to \infty } } \right|^2
\\
   = \frac{{\hbar ^2 }}
{{d^2 }}\frac{{\Gamma ^2 }} {4}g^2 (\omega _0 )\Delta \left|
{\frac{{e^{ - i2\sqrt \Lambda  \left( {t + \frac{x} {{v_g }}}
\right)} }} {{\omega _0  - \Omega  + 2\sqrt \Lambda   + i\Gamma }}
+ \frac{{e^{ - i2\sqrt \Lambda  \left( {t + \frac{x} {{v_g }}}
\right)} }}
{{\omega _0  - \Omega  - 2\sqrt \Lambda   + i\Gamma }}} \right|^2
\\
+ \frac{{\hbar ^2 }}
{{d^2 }}\frac{{\Gamma ^2 }} {{16}}g^2 (\omega _0 )\Delta \left|
{\left( {\frac{{e^{ - i2\sqrt \Lambda  \left( {t + \frac{x} {{v_g
}}} \right)} }} {{\omega _0  - \Omega  + 2\sqrt \Lambda   +
i\Gamma }} - \frac{{e^{i2\sqrt \Lambda  \left( {t + \frac{x} {{v_g
}}} \right)} }}
{{\omega _0  - \Omega  - 2\sqrt \Lambda   + i\Gamma }}} \right)} \right|^2.
\end{gathered}
   \end{equation}
\end{widetext}

The expressions (\ref{4.25}), (\ref{4.26}) are the probability
densities, measured by a detector located at the point $x$. In what
follows we will use for $g(\omega_0)$ the resonance approximation,
$g(\omega_0) \approx g(\Omega)=\sqrt{\Gamma/2\pi}$.

As is seen from (\ref{4.25}), (\ref{4.26}) the frequency spectrum
of the first-order correlation function exhibits oscillations with
low frequency $4\sqrt{\Lambda}$. After averaging over period of
these oscillations $G_{av}^{(1)} = \frac{1}{T}\int\limits_0^T
{G^{(1)} (x,t)dt}$, where $T=2\pi/4\sqrt{\Lambda}$,
 we obtain the
stationary frequency spectrum:
  \begin{subequations}
  \begin{align}
\begin{gathered}
g_{av,x>0}^{(1)} = 1 + \frac{5}{16} \frac{\Gamma^2}{(\omega_0 -
\Omega +2 \sqrt{\Lambda})^2 + \Gamma^2}
\\
+ \frac{5}{16} \frac{\Gamma^2}{(\omega_0 - \Omega -2 \sqrt{\Lambda})^2 + \Gamma^2},
\end{gathered} \label{g1_norm_posit}
\\
\begin{gathered}
g_{av,x<0}^{(1)} = \frac{5}{16} \frac{\Gamma^2}{(\omega_0 - \Omega
+2 \sqrt{\Lambda})^2 + \Gamma^2}
\\
+ \frac{5}{16} \frac{\Gamma^2}{(\omega_0 - \Omega -2 \sqrt{\Lambda})^2 + \Gamma^2},
\end{gathered}   \label{g1_norm_negat}
  \end{align}
  \end{subequations}
Here we normalize correlations functions (\ref{4.25}-\ref{4.26})
to the squared amplitude of incoming wave $ g_{av}^{(1)}  =
G_{av}^{(1)}\frac{{d^2 }} {{\hbar ^2 \Delta g^2 (\omega _0
)}}$

\begin{figure}
\includegraphics[width=8 cm]{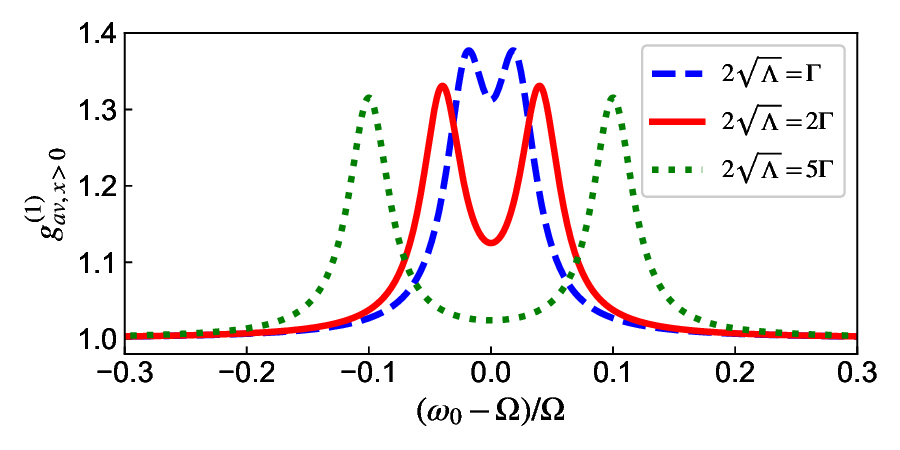}\\
  \caption{Frequency spectra of the normalized first-order correlation
  function (\ref{g1_norm_posit}) for transmitted field, $x >0$. Different lines correspond
  to different frequency shifts: blue
dashed line, $2\sqrt{\Lambda} = \Gamma $; red solid line,
$2\sqrt{\Lambda} = 2\Gamma $; Green dotted line, $2\sqrt{\Lambda}
= 5\Gamma $; Other parameters: $\Gamma /\Omega = 0.02, \;
\Delta/\Omega = 0.1$.}  \label{fig1}
  \end{figure}

The dependence of stationary correlation function
(\ref{g1_norm_posit}) is shown in Fig. \ref{fig1} for different
values of frequency shift $2\sqrt{\Lambda}$. Every plot exhibits
two emission peaks with the distance $4\sqrt{\Lambda}$ between
them. These peaks are the signature of the stimulated Rabi
oscillations in open waveguide between ground and excites states
of a two-level atom with the frequency $2\sqrt{\Lambda}$.

\subsection{Second-order correlation function $G^{(2)}(x_2,t_2;x_1,t_1)$}

For second-order correlation function (\ref{2.7}) we need four
matrix elements, $\left\langle g,0|E^+(x_2,t_2)|e,0\right\rangle$,
$\left\langle e,0|E^+(x_1,t_1)|E,1\right\rangle$, $\left\langle
g,0|a(\omega,0)E^+(x_1,t_1)|E,1\right\rangle$, and $\left\langle
g,0|E^+(x_2,t_2)a^\dagger (\omega,0)|g,0\right\rangle$. First three
matrix elements were given above in the expressions (\ref{4.1}),
(\ref{4.15}), and (\ref{4.16}), respectively. The matrix element
$\left\langle g,0|E^+(x_2,t_2)a^\dagger (\omega,0)|g,0\right\rangle$ is
as follows:
  \begin{equation}\label{g3}
\begin{gathered}
  \left\langle {g,0} \right|E^ +  (x_2 ,t_2 )a^\dagger  (\omega ')\left| {g,0} \right\rangle  = i\frac{\hbar }
{d}g(\omega ')e^{ - i\omega '\left( {t_2  - \frac{{x_2 }}
{{v_g }}} \right)}
\\
   - i \frac{\hbar }{d} \Gamma e^{ - i(\Omega  - i\Gamma )(t_2  - \frac{{x_2 }}
{{v_g }})} \theta (x_2 ) g(\omega ')\int\limits_0^{t_2  - \frac{{x_2 }}
{{v_g }}} {d\tau } e^{ - i(\omega ' - \Omega  + i\Gamma )\tau }
\\
   - i \frac{\hbar }
{d}\Gamma e^{ - i(\Omega  - i\Gamma (t_2  + \frac{{x_2 }}
{{v_g }})} \theta ( - x_2 ) g(\omega ')\int\limits_0^{t_2  + \frac{{x_2 }}
{{v_g }}} {d\tau } e^{ - i(\omega ' - \Omega  + i\Gamma )\tau }.
\end{gathered}
  \end{equation}
After integration over $\tau$ we obtain:
   \begin{equation}\label{g2}
\begin{gathered}
\left\langle {g,0} \right|E^ +  (x_2 ,t_2 )a^\dagger  (\omega ')\left|
{g,0} \right\rangle  = i\frac{\hbar } {d}g(\omega ')e^{ - i\omega
'\left( {t_2  - \frac{{x_2 }}
{{v_g }}} \right)}
\\
   + \frac{\hbar }
{d}\Gamma g(\omega ')\frac{{e^{ - i\omega '\left( {t_2  -
\frac{{x_2 }} {{v_g }}} \right)}  - e^{ - i(\Omega  - i\Gamma
)\left( {t_2  - \frac{{x_2 }} {{v_g }}} \right)} }} {{\omega ' -
\Omega  + i\Gamma }}
\\
\times\theta (x_2 )\theta \left( {t_2  - \frac{{x_2 }}
{{v_g }}} \right)
\\
   + \frac{\hbar }
{d}\Gamma g(\omega ')\frac{{e^{ - i\omega '\left( {t_2  +
\frac{{x_2 }} {{v_g }}} \right)}  - e^{ - i(\Omega  - i\Gamma
)\left( {t_2  + \frac{{x_2 }} {{v_g }}} \right)} }} {{\omega ' -
\Omega  + i\Gamma }}
\\
\times\theta ( - x_2 )\theta \left( {t + \frac{{x_2 }}
{{v_g }}} \right).
\end{gathered}
   \end{equation}

Below we use plain wave approximation to write these matrix
elements for different space locations of the detectors. The
corresponding expressions are given in Appendix D.

\section{Photon correlations - The Hanbury Brown and Twiss effect for radiation from a two-level atom}

Here we study the dependence of the correlation between two
measurements on the delay time $\Delta t=t_2-t_1$. Note, that
always $t_2\geq t_1$ since a second detector measures the field
$\Delta t$ seconds later. For this reason we assume $|x_2|>|x_1|
$, that is, the signal first reaches the first detector. As the
operators $E^+(x_1,t_1)$ and $E^+(x_2,t_2)$ do not commute, it is
important to distinguish between two detectors. The first
measurement changes the states of the field. We may say that after
the first measurement the two-excitation state of the field
$|E,1\rangle$ jumps to the superposition of the single-excitation
states $|e,0\rangle$ and $|g,1\rangle$. Therefore, the second
detector measures the field which differs from the field measured
earlier by the first detector. That is the reason why the
off-diagonal matrix elements of $E^+(x_1,t_1)$ do not coincide
with those of $E^+(x_2,t_2)$.

For convenience we introduce new time variables, $T_1=t_1-
|x_1|/v_g$, $T_2=t_2-|x_2|/v_g$. We count the time from the moment
the signal reaches the first detector, $t^{(0)}_1=|x_1|/v_g$. The
measurement at the second detector is performed some time later at
$t_2=t^{(0)}_2+\Delta T$, where $t^{(0)}_2$ is the time it takes
for a signal to reach the second detector,
$t^{(0)}_2=t^{(0)}_1+\frac{|x_2|-|x_1|}{v_g}$. Therefore,
$t_2-\frac{|x_2|}{v_g}=t^{(0)}_1-\frac{|x_1|}{v_g}+\Delta T$.
Thus, at the initial moment $T_1=0$, $T_2=\Delta T$.

   \begin{figure*}
\includegraphics[width=17 cm]{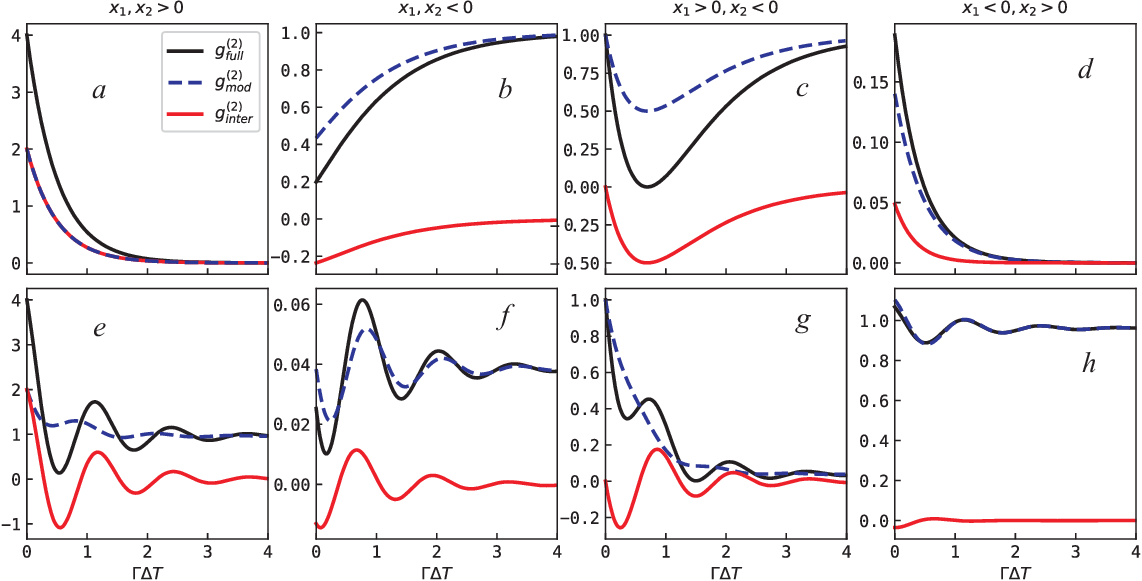}\\
\caption{The dependence of second-order correlation function on
delay time $\Delta T$ for different positions of detectors (shown
at the top of each plot). Blue dashed line corresponds to the sum
of bare squared modulus of the correlation function
$g^{(2)}_{mod}=(G^{(2)}_{path1}+G^{(2)}_{path2})2\pi
d^4/\hbar^4\Delta\Gamma^3$; red line corresponds to the
interference term $g^{(2)}_{int}=G^{(2)}_{int}2\pi
d^4/\hbar^4\Delta\Gamma^3$; black solid line corresponds to full
correlation function $g^{(2)}_{full}=G^{(2)}_{full}2\pi
d^4/\hbar^4\Delta\Gamma^3$. Top row of graphs (a)-(d) is plotted
for resonant frequency of incident field, $\omega_0/\Omega = 1$.
Bottom row (e)-(h) is plotted for detuned incident field,
$\omega_0/\Omega = 1.1$. Decay rate $\Gamma /\Omega =
0.02$.}\label{fig2}
   \end{figure*}

First we take $x_1$ and $x_2$ on the positive $x$-axis, so that
$x_1<x_2$. Using (\ref{4.27a}), (\ref{4.28a}), (\ref{4.29a}),
(\ref{4.30a}) and substituting $t_2 - x/\upsilon_g = \Delta T$, we
construct the second-order correlation function (\ref{2.7}) for
two detectors positioned at the positive $x$-axis:
   \begin{equation}\label{7.2}
\begin{gathered}
  G^{(2)} (x_2 ,t_2 ;x_1 ,t_1 )_{_{x_2  > 0}^{x_1  > 0} }  = \frac{{\hbar ^4 }}
{{d^4 }}\Gamma ^2 \Delta g^2 (\omega_0 )
\\
   \times \left| {1 - i\Gamma \frac{{1 - e^{ - i(\Omega  - \omega _0  - i\Gamma )\Delta T} }}
{{\omega _0  - \Omega  + i\Gamma }} + e^{ - i(\Omega  - \omega _0  - i\Gamma )\Delta T} } \right|^2 .
\end{gathered}
   \end{equation}

For two detectors positioned at negative $x$-axis at $x_1<0$ and
$x_2<0$ we use (\ref{4.27b}), (\ref{4.28b}), (\ref{4.29b}),
(\ref{4.30b}) to obtain the second-order correlation function:
\begin{widetext}
   \begin{equation}\label{7.4}
\begin{gathered}
  G^{(2)} (x_2 ,t_2 ;x_1 ,t_1 )_{_{x_2 < 0}^{x_1 < 0} }  = \frac{{\hbar ^4 }}
{{d^4 }}\Gamma ^2 g^2 (\omega _0 )\Delta \left| {\Gamma \frac{{1 -
e^{ - i(\Omega  - \omega _0  - i\Gamma )\Delta T} }}
{{\omega _0  - \Omega  + i\Gamma }}} \right.
\\
   - \frac{\Gamma }
{2}\left. {\left( {\frac{{e^{ - i2\sqrt \Lambda  \Delta t_1 }  -
e^{ - i(\Omega  - \omega _0  - i\Gamma )\Delta t_1 } }} {{\omega
_0  - \Omega  + 2\sqrt \Lambda   + i\Gamma }} + \frac{{e^{i2\sqrt
\Lambda  \Delta t_1 }  - e^{ - i(\Omega  - \omega _0  - i\Gamma
)\Delta t_1 } }}
{{\omega _0  - \Omega  - 2\sqrt \Lambda   + i\Gamma }}} \right)e^{ - i(\Omega  - \omega _0  - i\Gamma )\Delta T} } \right|^2.
\end{gathered}
   \end{equation}

If the first detector is at $x_1>0$ and the second detector is at
$x_2<0$, we obtain from (\ref{4.27b}), (\ref{4.28a}),
(\ref{4.29a}), (\ref{4.30b}):
   \begin{equation}\label{7.5}
\begin{gathered}
  G^{(2)} (x_2 ,t_2 ;x_1 ,t_1 )_{_{x_2  < 0}^{x_1  > 0} } =
\frac{{\hbar ^4 }}
{{d^4 }}\Gamma ^2 \Delta g^2 (\omega _0 )\left| {i\Gamma \frac{{1
- e^{ - i(\Omega  - \omega _0  - i\Gamma )\Delta T} }}
{{\omega _0  - \Omega  + i\Gamma }} - e^{ - i(\Omega  - \omega _0  - i\Gamma )\Delta T} } \right|^2  .
\end{gathered}
   \end{equation}

For the first detector at $x_1<0$, and the second detector at
$x_2>0$, we obtain from (\ref{4.27a}), (\ref{4.28b}),
(\ref{4.29b}), (\ref{4.30a}):
   \begin{equation}\label{7.6}
\begin{gathered}
  G^{(2)} (x_2 ,t_2 ;x_1 ,t_1 )_{_{x_2 > 0}^{x_1 < 0} }  = \frac{{\hbar ^4 }}
{{d^4 }}\Gamma ^2 \Delta g^2 \left( {\omega _0 } \right)\left|
{\left( {1 - i\Gamma \frac{{1 - e^{ - i(\Omega  - \omega _0  -
i\Gamma )\Delta T} }}
{{\omega _0  - \Omega  + i\Gamma }}} \right)} \right.
\\
   + \left. {ie^{ - i(\Omega  - \omega _0  - i\Gamma )\Delta T} e^{ - i\omega _0 \Delta t_1 } \frac{\Gamma }
{2}\left( {\frac{{e^{ - i2\sqrt \Lambda  \Delta t_1 }  - e^{ -
i(\Omega  - \omega _0  - i\Gamma )\Delta t_1 } }} {{\omega _0  -
\Omega  + 2\sqrt \Lambda   + i\Gamma }} + \frac{{e^{i2\sqrt
\Lambda  \Delta t_1 }  - e^{ - i(\Omega  - \omega _0  - i\Gamma
)\Delta t_1 } }}
{{\omega _0  - \Omega  - 2\sqrt \Lambda   + i\Gamma }}} \right)} \right|^2.
\end{gathered}
   \end{equation}
   \end{widetext}

The dependence of second-order correlation function on delay time
$\Delta T$ for different positions of detectors is shown in Fig.
\ref{fig2}. In this figure we plot the dimensionless normalized
quantity $g^{(2)}(x,t)=G^{(2)}2\pi d^4/\hbar^4 \Delta\Gamma^3$.
Plots (\emph{a}) and (\emph{e}) are calculated from the expression
(\ref{7.2}); plots (\emph{b}), (\emph{f}) are calculated from
(\ref{7.4}); plots (\emph{c}), (\emph{g}) are calculated from
(\ref{7.5}) and plots (\emph{d}), (\emph{f}) are calculated from
(\ref{7.6}). The contribution of the sum
$g^{(2)}_{mod}=(G^{(2)}_{path1}+G^{(2)}_{path2})2\pi
d^4/\hbar^4\Delta\Gamma^3$ is shown by blue dashed line. The
contribution of the interference term (\ref{2.10}),
$g^{(2)}_{int}=G^{(2)}_{int}2\pi d^4/\hbar^4\Delta\Gamma^3$ is
shown by the red solid line. The full second-order correlation
function (\ref{2.7}), which accounts for the interference
contribution, is shown by the black solid line. Plots
Fig.\ref{fig2} \emph{a}, \emph{b}, \emph{c}, \emph{d} are
calculated for resonant incident photon, $\omega_0=\Omega$.

We note that in equations (\ref{7.4}), (\ref{7.6}) the second
terms, which provide the interference in Fig. \ref{fig2}\emph{b},
\emph{e}, \emph{d}, \emph{h}, come from the matrix element
(\ref{4.28b}), $\langle e,0|E^+(x_1,t_1|E,1\rangle$ for $x_1<0$.
At the time $t_1=-x_1/v_g$ this matrix element equals zero.  The
detector absorbs non-zero signal at the later time
$t=-x_1/v_g+\Delta t_1$, where $\Delta t_1$ is on the order of
$1/\Gamma$. Therefore, in order to preserve the interference for
the cases shown in Fig. \ref{fig2}\emph{b}, \emph{e}, \emph{d},
\emph{h}, we begin to count delay time $\Delta T$ relative to
$\Delta t_1$, where we take $\Delta t_1=1/\Gamma$.

From these results we see that the type of interference
(constructive or destructive) depends crucially on the mutual
positions of two detectors. It is understandable because the
measurements of two detectors are not independent: the first
detector measure the field in the initial state $|e,1\rangle$,
while the second detector measures the field which is in the
superposition of the one-excitation states $|g,1\rangle$ and
$|e,0\rangle$.

The constructive interference is clearly seen in
Fig.\ref{fig2}\emph{a} on resonance, $\omega_0=\Omega$, when two
detectors are on the positive x-axis. It increases the probability
density above the level of bare contribution $g^{(2)}_{mod}$. In
this case, the square modulus in (\ref{7.2}) is equal to
$4e^{-2\Gamma\Delta T}$. The factor $4$ comes from the
interference contribution. Without the interference we would have
$2e^{-2\Gamma\Delta T}$. The constructive interference is also
seen if the first detector is on negative x-axis and the second
detector is on the positive x-axis, Fig.\ref{fig2}\emph{d}.

The destructive interference is observed in
Fig.\ref{fig2}\emph{b}, \emph{c}. On these plots the interference
term is negative. It decreases the probability density below the
level of bare contribution $g^{(2)}_{mod}$. We note that for
Fig.\ref{fig2} \emph{c} the total probability density is exactly
equal to zero at the point $\Gamma\Delta T= \ln 2$, that follows
from (\ref{7.5}).

For relatively small detuning $|\omega_0-\Omega|\ll\Omega$, the
contribution of interference may be constructive or destructive
depending on the $\Delta T$ (plots (\emph{e}) and (\emph{g})). For
$\Delta T\rightarrow\infty$ the interference disappears.

\section{Conclusions}

In this paper we consider a single-photon scattering on a
two-level atom embedded in an open waveguide. We show that the
scattering photon probability amplitudes are given by off-diagonal
matrix elements of the quantum operator of electric field, which
accounts for the interaction  of atom with continuum of electric
field modes in an open waveguide. For a one- and two-excitation
subspace we find the first-order and second-order photon
correlation functions which describe the photon probability
densities when the field is measured by one or two detectors,
correspondingly. We show that the second order correlation
function describes the interference between the measurements by
two detectors. This interference, which arises due to two possible
pathways of two photons (one scattered photon and other
spontaneously emitted photon) is similar to the well known Hanbury
Brown and Twiss interference effect. We show that the interference
may be constructive or destructive, which strongly depends on the
positions of two detectors relative to the origin of coordinate,
$x=0$ where the two-level atom is located.

Final expressions for correlation functions in this work were calculated
for incident single-photon plane wave. However, general expressions explicitly
contain pulse shape function $f(\omega )$ which allows one to use any arbitrary single-photon pulse shape, for example Gaussian packet.

\begin{acknowledgments}

The authors thank O. V. Kibis for fruitful discussions. The work
is supported by the Ministry of Science and Higher Education of
Russian Federation under the project FSUN-2023-0006. O. Chuikin
acknowledges the financial support from the Foundation for the
Advancement of Theoretical Physics and Mathematics “BASIS”.

\end{acknowledgments}

\appendix

\section{Derivation of the expression (\ref{10})}

For $\sigma_-(t)$ the Heisenberg equation of motion reads:
   \begin{equation}\label{A1}
\begin{gathered}
\frac{{d\sigma_- }}
{{dt}} =  - i\Omega \sigma_-  (t)
   + i\int\limits_0^\infty  {g\left( \omega  \right)\left( {\sigma _Z (t)a(\omega ,t)} \right)d\omega }
\\
   + i\int\limits_0^\infty  {g\left( \omega  \right)\left( {\sigma _Z (t)b(\omega ,t)} \right)d\omega }.
\end{gathered}
   \end{equation}
The transformation to "slow" variable $\sigma_-(t)\rightarrow
{\tilde{\sigma}}_-(t)e^{-i\Omega t}$ transforms (\ref{A1}) to
   \begin{equation}\label{A2}
\begin{gathered}
  \frac{{d\tilde{\sigma}_ -  }}
{{dt}} =  + i\int\limits_0^\infty  {g\left( \omega  \right)\sigma
_Z (t)e^{ i\Omega t} a(\omega ,t)d\omega }
\\
   + i\int\limits_0^\infty  {g\left( \omega  \right)\sigma _Z (t)e^{ i\Omega t} b(\omega ,t)d\omega } .
\end{gathered}
   \end{equation}

Using the expressions (\ref{a}) and (\ref{b}) for photon operators
in (\ref{A2}) we obtain:
   \begin{equation}\label{A3}
\begin{gathered}
  \frac{{d\tilde{\sigma}_-}  }
{{dt}} = i\sigma _Z (t)\int\limits_0^\infty  {g\left( \omega  \right)e^{ - i(\omega  - \Omega )t} a(\omega ,0)d\omega }
\\
   + i\sigma _Z (t)\int\limits_0^\infty  {g\left( \omega  \right)e^{ - i(\omega  - \Omega )t} b(\omega ,0)d\omega }
\\
+ 2\int\limits_0^\infty  {g^2 (\omega )\left( {\int\limits_0^t {e^{ - i(\omega  - \Omega )(t - \tau )} \sigma_Z (t)\tilde{\sigma}_-  (\tau )d\tau } } \right)d\omega } .
\end{gathered}
   \end{equation}
In order to calculate the integral $\int\limits_0^t {e^{ -
i(\omega - \Omega )(t - \tau )} \sigma _Z (t)\tilde{\sigma} _ -
(\tau )d\tau }$ we assume
$\tilde{\sigma}_-(\tau)=\tilde{\sigma}_-(t)$. This assumption is
equivalent to Wigner-Weisskopf or Markov approximations. It
allows us to take $\tilde{\sigma}_-(\tau)$ out of the integrand in
the last line of (\ref{A3}). Therefore, we obtain:
   \begin{equation}\label{A4}
\begin{gathered}
  \int\limits_0^t {e^{ - i(\omega  - \Omega )(t - \tau )} \sigma _Z (t) \tilde \sigma _-  (\tau )d\tau }
\\
   \approx \sigma _Z (t)\tilde \sigma _ -  (t)\int\limits_0^\infty {e^{ - i(\omega  - \Omega )(t - \tau )} d\tau }
\\
   =  - \tilde \sigma_-  (t)\left( {\pi \delta (\omega  - \Omega ) - iPv\left( {\frac{1}
{{\omega  - \Omega }}} \right)} \right) ,
\end{gathered}
   \end{equation}
where we used the known property of the spin operators,
$\sigma_Z\sigma_-=-\sigma_-$.
The principal value $Pv(1/(\omega-\Omega))$ shifts the qubit
frequency $\Omega$. Assuming this shift is small, we include it
implicitly in renormalized qubit frequency.

Therefore, disregarding principal value we obtain:
   \begin{equation}\label{A5}
\int\limits_0^t {e^{ - i(\omega  - \Omega )(t - \tau )} \sigma _Z
(t)\tilde \sigma _ -  (\tau )d\tau }  \approx  - \tilde \sigma _ -
(t)\pi \delta (\omega  - \Omega ).
   \end{equation}
Under these approximations, the equation (\ref{A3}) takes the form:
   \begin{equation}\label{A6}
\begin{gathered}
  \frac{{d\tilde{\sigma} _ -  }}
{{dt}} = i\sigma _Z (t)\int\limits_0^\infty  {g\left( \omega  \right)e^{ - i(\omega  - \Omega )t} a(\omega ,0)d\omega }
\\
   + i\sigma _Z (t)\int\limits_0^\infty  {g\left( \omega  \right)e^{ - i(\omega  - \Omega )t} b(\omega ,0)d\omega }  - \Gamma \tilde{\sigma} _ -  (t),
\end{gathered}
   \end{equation}
where $\Gamma=2\pi g^2(\Omega)$.

The formal solution of equation (\ref{A6}) reads
   \begin{equation}\label{A7}
\begin{gathered}
  \tilde \sigma_- (t) = \sigma_- (0)e^{ - \Gamma t}
\\
 +i\int\limits_0^\infty  {d\omega } g(\omega )\int\limits_0^t {d\tau } e^{ - i(\omega  - \Omega )\tau } e^{ - \Gamma (t - \tau )} \sigma _Z (\tau )a(\omega ,0)
\\
   + i\int\limits_0^\infty  {d\omega } g(\omega )\int\limits_0^t {d\tau } e^{ - i(\omega  - \Omega )\tau } e^{ - \Gamma (t - \tau )} \sigma _Z (\tau )b(\omega ,0).
\end{gathered}
   \end{equation}
Coming back to $\sigma_-(t)$ ($\tilde{\sigma}_-(t)=e^{i\Omega
t}\sigma_-(t)$), we obtain equation (\ref{10}) from the main text.

\section{Derivation of the expression (\ref{4.3})}

The action of $E^+(x,t)$ on $\left|G,1\right\rangle$ is as
follows:
\begin{widetext}
   \begin{equation}\label{B1}
\begin{gathered}
  E^ +  (x,t)\left| {G,1} \right\rangle  = i\frac{\hbar }
{d}\int\limits_0^\infty  {d\omega } g(\omega )f(\omega )e^{ -
i\omega \left( {t - \frac{x}
{{v_g }}} \right)} \left| {g,0} \right\rangle
\\
   + \frac{\hbar }
{d}\Gamma e^{ - i\Omega (t - \frac{x} {{v_g }})} \theta
(x)i\int\limits_0^\infty  {d\omega } g(\omega )f(\omega
)\int\limits_0^{t - \frac{x} {{v_g }}} {d\tau } e^{ - i(\omega  -
\Omega )\tau } e^{ - \Gamma (t - \frac{x}
{{v_g }} - \tau )} \sigma _Z (\tau )\left| {g,0} \right\rangle
\\
   + \frac{\hbar }
{d}\Gamma e^{ - i\Omega (t + \frac{x} {{v_g }})} \theta ( -
x)i\int\limits_0^\infty  {d\omega } g(\omega )f(\omega
)\int\limits_0^{t + \frac{x} {{v_g }}} {d\tau } e^{ - i(\omega  -
\Omega )\tau } e^{ - \Gamma (t + \frac{x}
{{v_g }} - \tau )} \sigma _Z (\tau )\left| {g,0} \right\rangle.
\end{gathered}
   \end{equation}
The calculation of the quantity
$\sigma_Z(\tau)\left|g,0\right\rangle$ is straightforward. As
$[\sigma_Z,H_0]=0$,
$\sigma_Z(\tau)=e^{iH_{JC}\tau}\sigma_Z(0)e^{-iH_{JC}{\tau}}$,
where $H_{JC}$ is the Jaynes-Cummings Hamiltonian (\ref{1b}).
Since $H_{JC}\left|g,0\right\rangle=0$, we have $e^{\pm i
H_{JC}\tau}\left|g,0\right\rangle=\left|g,0\right\rangle$. Thus,
$\sigma_Z(\tau)\left|g,0\right\rangle=-\left|g,0\right\rangle$,
where negative sign is due to
$\sigma_Z(0)\left|g\right\rangle=-\left|g\right\rangle$.
Therefore, there is the only one non-zero matrix element:
   \begin{equation}\label{B2}
\begin{gathered}
  \left\langle {g,0} \right|E^ +  (x,t)\left| {G,1} \right\rangle  = i\frac{\hbar }
{d}\int\limits_0^\infty  {d\omega } g(\omega )f(\omega )e^{ -
i\omega \left( {t - \frac{x}
{v_g }} \right)}
\\
   - \frac{\hbar }
{d}\Gamma e^{ - i\Omega (t - \frac{x} {{v_g }})} \theta
(x)i\int\limits_0^\infty  {d\omega } g(\omega )f(\omega
)\int\limits_0^{t - \frac{x} {{v_g }}} {d\tau } e^{ - i(\omega  -
\Omega )\tau } e^{ - \Gamma (t - \frac{x}
{{v_g }} - \tau )}
\\
   - \frac{\hbar }
{d}\Gamma e^{ - i\Omega (t + \frac{x} {{v_g }})} \theta ( -
x)i\int\limits_0^\infty  {d\omega } g(\omega )f(\omega
)\int\limits_0^{t + \frac{x} {{v_g }}} {d\tau } e^{ - i(\omega  -
\Omega )\tau } e^{ - \Gamma (t + \frac{x}
{{v_g }} - \tau )}.
\end{gathered}
    \end{equation}
Calculating the integrals over $\tau$, we obtain the expression
(\ref{4.3}) in the main text.

\section{Derivation of the expression (\ref{4.9})}

First we write down the action of $E^+(x,t)$ on the initial state
$\left|E,1\right\rangle$
   \begin{equation}\label{C1}
\begin{gathered}
  E^ +  (x,t)\left| {E,1} \right\rangle  = i\frac{\hbar }
{d}\int\limits_0^\infty  {d\omega } g(\omega )f(\omega )e^{ -
i\omega \left( {t - \frac{x}
{{v_g }}} \right)} \left| {e,0} \right\rangle
\\
   + \frac{\hbar }
{d}\Gamma e^{ - i\Omega (t - \frac{x} {{v_g }})} \theta
(x)i\int\limits_0^\infty  {d\omega } g(\omega )f(\omega
)\int\limits_0^{t - \frac{x} {{v_g }}} {d\tau } e^{ - i(\omega  -
\Omega )\tau } e^{ - \Gamma (t - \frac{x}
{{v_g }} - \tau )} \sigma _Z (\tau )\left| {e,0} \right\rangle
\\
   + \frac{\hbar }
{d}\Gamma e^{ - i\Omega (t + \frac{x} {{v_g }})} \theta ( -
x)i\int\limits_0^\infty  {d\omega } g(\omega )f(\omega
)\int\limits_0^{t + \frac{x} {{v_g }}} {d\tau } e^{ - i(\omega  -
\Omega )\tau } e^{ - \Gamma (t + \frac{x}
{{v_g }} - \tau )} \sigma _Z (\tau )\left| {e,0} \right\rangle
\\
   + \frac{\hbar }
{d}\Gamma \theta ( - x)e^{ - i(\Omega  - i\Gamma )(t + \frac{x}
{{v_g }})} \left| {G,1} \right\rangle  + \frac{\hbar } {d}\Gamma
\theta (x)e^{ - i(\Omega  - i\Gamma )(t - \frac{x} {{v_g }})}
\left| {G,1} \right\rangle.
\end{gathered}
   \end{equation}
\end{widetext}
The key point here is the calculation  of the quantity
$\sigma_Z(\tau)\left|e,0\right\rangle$. As $[\sigma_Z,H_0]=0$,
$\sigma_Z(\tau)=e^{iH_{JC}\tau}\sigma_Z(0)e^{-iH_{JC}{\tau}}$,
where $H_{JC}$ is the Jaynes-Cummings Hamiltonian (\ref{1b}).

First we prove the following relation:
   \begin{equation}\label{C14}
\sigma _Z (0)e^{ - iH_{JC} \tau }  = e^{iH_{JC} \tau } \sigma _Z
(0).
   \end{equation}
From the series expansion of exponents we see that
(\ref{C14}) is valid, if the following relations holds:
   \begin{equation}\label{C15}
( - 1)^n \sigma _Z (0)H_{JC}^n  = H_{JC}^n \sigma _Z (0),{\text{
}}n = 1,2,3....
   \end{equation}
Using the properties of Pauli operators $\sigma_Z, \sigma_+,
\sigma_-$, it is not difficult to show that the equation
(\ref{C14}) indeed holds for the Jaynes-Cummings Hamiltonian
(\ref{1b}). Therefore,
$\sigma_Z(\tau)\left|e,0\right\rangle=e^{iH_{JC}2\tau}\left|e,0\right\rangle$.

Direct calculations show that:
   \begin{equation}\label{C2}
H_{_{JC} }^{2n + 1} \left| {e,0 } \right\rangle  = \Lambda^n
\left| \overline{{G,1 }} \right\rangle ,\quad n = 0,1,2,3.....
   \end{equation}
   \begin{equation}\label{C3}
H_{_{JC} }^{2n} \left| {e,0 } \right\rangle  = \Lambda^n \left|
{e,0 } \right\rangle ,\quad n = 1,2,3.....
   \end{equation}
where
   \begin{equation}\label{C4}
\overline {\left| {G,1} \right\rangle }  = \int\limits_0^\infty
{d\omega } g(\omega )a^\dagger  (\omega ,0)\left| {g,0}
\right\rangle,
   \end{equation}
   \begin{equation}\label{C5}
\Lambda  =  {\int\limits_0^\infty  {d\omega } g^2 (\omega )}.
   \end{equation}

From the series expansion of the exponent we obtain:
   \begin{equation}\label{C6}
\begin{gathered}
  e^{iH_{JC} 2\tau } \left| {e,0 } \right\rangle  = \sum\limits_{n = 0}^\infty  {} \frac{{\left( {iH_{JC} 2\tau } \right)^n }}
{{n{\text{!}}}}\left| {e,0 } \right\rangle
\\
   = \sum\limits_{n = 0}^\infty  {} \frac{{\left( {i2\tau } \right)^{2n} }}
{{\left( {2n} \right){\text{!}}}}\Lambda ^n \left| {e,0 }
\right\rangle  + \sum\limits_{n = 0}^\infty  {} \frac{{\left(
{i2\tau } \right)^{2n + 1} }} {{\left( {2n + 1}
\right){\text{!}}}}\Lambda ^n \left| {\overline {G,1 } }
\right\rangle .
\end{gathered}
   \end{equation}
The expression (\ref{C6}) can be rewritten as:
   \begin{equation}\label{C13}
\begin{gathered}
  e^{iH_{JC} \tau } \sigma _Z (0)e^{ - iH_{JC} \tau } \left| {e,0} \right\rangle
\\
   = \cos \left( {2\sqrt \Lambda  \tau } \right)\left| {e,0} \right\rangle  + \frac{i}
{{2\sqrt \Lambda  }}\sin \left( {2\sqrt \Lambda  \tau }
\right)\overline {\left| {G,1} \right\rangle }.
\end{gathered}
   \end{equation}
Finally, using (\ref{C13}) in (\ref{C1}), we obtain the expression
(\ref{4.9}).

\section{The expressions of matrix elements (\ref{4.1}), (\ref{4.15}), (\ref{4.16}), and (\ref{g2}) for different space locations of the detectors}

\begin{widetext}
\begin{subequations}\label{4.27}
    \begin{align}
\left\langle {g,0} \right|E^ +  (x_2 ,t_2 )\left| {e,0}
\right\rangle _{x_2  > 0}  = \frac{\hbar } {d}\Gamma e^{ -
i(\Omega  - i\Gamma )(t_2  - \frac{{x_2 }} {v_g })}, \label{4.27a}
\\
\left\langle {g,0} \right|E^ +  (x_2 ,t_2 )\left| {e,0}
\right\rangle _{x_2  < 0}  = \frac{\hbar } {d}\Gamma e^{ -
i(\Omega  - i\Gamma )(t_2  + \frac{{x_2 }} {v_g })}, \label{4.27b}
 \end{align}
    \end{subequations}

   \begin{subequations}\label{4.28}
  \begin{align}
  \begin{gathered}\label{4.28a}
\left\langle {e,0} \right|E^ +  (x_1 ,t_1 )\left| {E,1} \right\rangle _{x_1  > 0,t_1  > x_1 /v_g }  = i\frac{\hbar }
{d}g(\omega _0 )\sqrt \Delta  e^{ - i\omega _0 \left( {t_1  -
\frac{{x_1 }}
{{v_g }}} \right)}
\\
   \times \,\left( {1 + i\frac{\Gamma }
{2}\left( {\frac{{e^{i2\sqrt \Lambda  (\left( {t_1  - \frac{{x_1
}} {{v_g }}} \right)}  - e^{ - i(\Omega  - \omega _0  - i\Gamma
)\left( {t_1  - \frac{{x_1 }} {{v_g }}} \right)} }} {{\omega _0  -
\Omega  - 2\sqrt \Lambda   + i\Gamma }} + \frac{{e^{ - i2\sqrt
\Lambda  \left( {t_1  - \frac{{x_1 }} {{v_g }}} \right)}  - e^{ -
i(\Omega  - i\omega _0  - i\Gamma )\left( {t_1  - \frac{{x_1 }}
{{v_g }}} \right)} }}
{{\omega _0  - \Omega  + 2\sqrt \Lambda   + i\Gamma }}} \right)} \right),
  \end{gathered}
\\\nonumber
\\
  \begin{gathered}\label{4.28b}
  \left\langle {e,0} \right|E^ +  (x_1,t_1)\left| {E,1} \right\rangle _{x_1  < 0,t_1  >  - x_1 /v_g }  =  - \frac{\hbar }
{d}\frac{\Gamma } {2}g(\omega _0 )\sqrt \Delta  e^{ - i\omega _0
\left( {t_1  + \frac{{x_1 }}
{{v_g }}} \right)}
\\
   \times \left( {\frac{{e^{ - i2\sqrt \Lambda  \left( {t_1  + \frac{{x_1 }}
{{v_g }}} \right)}  - e^{ - i(\Omega  - \omega _0  - i\Gamma
)\left( {t_1  + \frac{{x_1 }} {{v_g }}} \right)} }} {{\omega _0  -
\Omega  + 2\sqrt \Lambda   + i\Gamma }} + \frac{{e^{i2\sqrt
\Lambda  \left( {t_1  + \frac{{x_1 }} {{v_g }}} \right)}  - e^{ -
i(\Omega  - \omega _0  - i\Gamma )\left( {t_1  + \frac{{x_1 }}
{{v_g }}} \right)} }}
{{\omega _0  - \Omega  - 2\sqrt \Lambda   + i\Gamma }}} \right),
  \end{gathered}
  \end{align}
  \end{subequations}
\begin{subequations}\label{4.29}
\begin{align}
\begin{gathered}\label{4.29a}
  \left\langle {g,0} \right|a(\omega ',0)E^ +  (x_1 ,t_1 )\left| {E,1} \right\rangle _{x_1  > 0,t > x_1 /v_g }  = \frac{\hbar }
{d}\Gamma \sqrt \Delta  \delta (\omega ' - \omega _0 )e^{i(\Omega
- i\Gamma )\left( {t_1  - \frac{{x_1 }}
{{v_g }}} \right)}
\\
   - \frac{\hbar }
{d}\frac{\Gamma } {2}\frac{{g(\omega ')g(\omega _0 )\sqrt \Delta
}} {{\sqrt \Lambda  }}\frac{{e^{ - i(\omega _0  - 2\sqrt \Lambda
)\left( {t_1  - \frac{{x_1 }} {{v_g }}} \right)}  - e^{ - i(\Omega
- i\Gamma )\left( {t_1  - \frac{{x_1 }} {{v_g }}} \right)} }}
{{\omega _0  - \Omega  - 2\sqrt \Lambda   + i\Gamma }}
\\
   + \frac{\hbar }
{d}\frac{\Gamma } {2}\frac{{g(\omega ')g(\omega _0 )\sqrt \Delta
}} {{\sqrt \Lambda  }}\frac{{e^{ - i(\omega _0  + 2\sqrt \Lambda
)\left( {t_1  - \frac{{x_2 }} {{v_g }}} \right)}  - e^{ - i(\Omega
- i\Gamma )\left( {t_1  - \frac{{x_1 }} {{v_g }}} \right)} }}
{{\omega _0  - \Omega  + 2\sqrt \Lambda   + i\Gamma }},
\end{gathered}
\\\nonumber
\\
\begin{gathered}\label{4.29b}
  \left\langle {g,0} \right|a(\omega ',0)E^ +  (x_1 ,t_1 )\left| {E,1} \right\rangle _{x_1  < 0,t_1  >  - x_1 /v_g }  = \frac{\hbar }
{d}\Gamma \sqrt \Delta  \delta (\omega ' - \omega _0 )e^{i(\Omega
- i\Gamma )\left( {t_1  + \frac{{x_1 }}
{{v_g }}} \right)}
\\
   + \frac{\hbar }
{d}\frac{\Gamma } {2}\frac{{g(\omega ')g(\omega _0 )\sqrt \Delta
}} {{\sqrt \Lambda  }}\left( {\frac{{e^{ - i(\omega  + 2\sqrt
\Lambda  )\left( {t_1  + \frac{{x_1 }} {{v_g }}} \right)} }}
{{\omega _0  - \Omega  + 2\sqrt \Lambda   + i\Gamma }} -
\frac{{e^{ - i(\omega  - 2\sqrt \Lambda  )\left( {t_1  +
\frac{{x_1 }} {{v_g }}} \right)} }}
{{\omega _0  - \Omega  - 2\sqrt \Lambda   + i\Gamma }}} \right),
\end{gathered}
\end{align}
\end{subequations}

\begin{subequations}\label{4.30}
\begin{align}
\begin{gathered}\label{4.30a}
\left\langle {g,0} \right|E^ +  (x_2 ,t_2 )a^\dagger  (\omega ')\left|
{g,0} \right\rangle _{x_2  > 0,t_2  > x_2 /v_g }  = i\frac{\hbar }
{d}g(\omega ')e^{ - i\omega '\left( {t_2  - \frac{{x_2 }} {{v_g
}}} \right)}  + \frac{\hbar } {d}\Gamma g(\omega ')\frac{{e^{ -
i\omega '\left( {t_2  - \frac{{x_2 }} {{v_g }}} \right)}  - e^{ -
i(\Omega  - i\Gamma )\left( {t_2  - \frac{{x_2 }} {{v_g }}}
\right)} }} {{\omega ' - \Omega  + i\Gamma }},
\end{gathered}
\\\nonumber
\\
\begin{gathered}\label{4.30b}
\left\langle {g,0} \right|E^ +  (x_2 ,t_2 )a^\dagger  (\omega ')\left|
{g,0} \right\rangle _{x_2  < 0,t_2  >  - x_2 /v_g }  = \frac{\hbar
} {d}\Gamma g(\omega ')\frac{{e^{ - i\omega '\left( {t_2  +
\frac{{x_2 }} {{v_g }}} \right)}  - e^{ - i(\Omega  - i\Gamma
)\left( {t_2  + \frac{{x_2 }} {{v_g }}} \right)} }} {{\omega ' -
\Omega  + i\Gamma }}.
\end{gathered}
\end{align}
\end{subequations}
\end{widetext}


\begin{thebibliography}{99}

\bibitem{Rai2001} J. M. Raimond, M. Brune, and S. Haroche, Manipulating
quantum entanglement with atoms and photons in a cavity, Rev. Mod.
Phys. \textbf{73}, 565 (2001)., 565 (2001).

\bibitem{Roy17}
D. Roy, C. M. Wilson, and O. Firstenberg, Strongly interacting
photons in one-dimensional continuum, Rev. Mod. Phys. \textbf{89},
021001 (2017).

\bibitem{Chang18}
D. E. Chang, J. S. Douglas, A. Gonzalez-Tudela, C.-L. Hung, and H.
J. Kimble, Quantum matter built from nanoscopic lattices of atoms
and photons, Rev. Mod. Phys. \textbf{90}, 031002 (2018).

\bibitem{Turschmann19}
P. T{\"u}rschmann, H. L. Jeannic, S. F. Simonsen, H. R. Haakh, S.
G{\"o}tzinger, V. Sandoghdar, P. Lodahl, and N. Rotenberg,
Coherent nonlinear optics of quantum emitters in nanophotonic
waveguides, Nanophotonics \textbf{8}, 1641 (2019).

\bibitem{Sheremet23}
A. S. Sheremet, M. I. Petrov, I. V. Iorsh, A. V. Poshakinskiy, and
A. N. Poddubny, Waveguide quantum electrodynamics: Collective
radiance and photon-photon correlations, Rev. Mod. Phys.
\textbf{95}, 015002 (2023).

\bibitem{Gu2017} X. Gu, A. F. Kockum, A. Miranowicz, Y.-X. Liu, and F. Nori,
Microwave photonics with superconducting quantum circuits, Phys.
Rep. 718, 1 (2017).

\bibitem{Ruos2017} J. Ruostekoski and J. Javanainen, Arrays of strongly coupled
atoms in a one-dimensional waveguide. Phys. Rev.  A  \textbf{96},
033857 (2017).

\bibitem{Lal2013} K. Lalumi$\grave{e}$re, B. C. Sanders, A. F. van Loo,
 A. Fedorov,  A. Wallraff,  and A. Blais, Input-output theory for waveguide QED with an ensemble
of inhomogeneous atoms. Phys. Rev.  A  \textbf{88}, 043806 (2013).

\bibitem{Chang2012} D. E. Chang, L. Jiang, A. V. Gorshkov, and H. J. Kimble,
Cavity QED with atomic mirrors. New J. Phys. \textbf{14}, 063003
(2012).

\bibitem{Mirho2019} M. Mirhosseini, E. Kim, X. Zhang, A. Sipahigil, P. B. Dieterle,
 A. J. Keller,A. Asenjo-Garcia, D. E. Chang, and O. Painter, Cavity quantum
electrodynamics with atom-like mirrors. Nature \textbf{569}, 692
(2019).

\bibitem{Brehm2021}J. D. Brehm, A. N. Poddubny, A. Stehli, T. Wolz, H. Rotzinger,
and A. V. Ustinov, Waveguide bandgap engineering with an array of
superconducting qubits, npj Quantum Materials  \textbf{6}, 10
(2021).

\bibitem{Loo2014}A. F. van Loo, A. Fedorov, K. Lalumi$\grave{e}$re, B. C. Sanders, A. Blais,
and A. Wallraff Photon-mediated interactions between distant
artificial atoms. Science \textbf{342}, 1494 (2014).

\bibitem{Shen2009} J.-T. Shen and S. Fan, Theory of single-photon transport in a single-mode waveguide. I. Coupling to a cavity containing a
two-level atom. Phys. Rev. A 79, 023837 (2009).

\bibitem{Cheng2017} M.-T. Cheng, J. Xu, and G. S. Agarwal,
Waveguide transport mediated by strong coupling with atoms. Phys.
Rev. A \textbf{95}, 053807 (2017).

\bibitem{Fang2014} Y.-L. L. Fang, H. Zheng, and H. U. Baranger,
One-dimensional waveguide coupled to multiple qubits photon-photon
correlations. EPJ Quantum Technol. \textbf{1}, 3 (2014).

\bibitem{Zheng2013} H. Zheng and H. U. Baranger, Persistent
Quantum Beats and Long-Distance Entanglement from
Waveguide-Mediated Interactions. Phys. Rev. Lett. \textbf{110},
113601 (2013).

\bibitem{Roy2011} D. Roy, Correlated few-photon transport in one-dimensional
waveguides: Linear and nonlinear dispersions, Phys. Rev. A
\textbf{83}, 043823 (2011).

\bibitem{Huang2013} J.-F. Huang, T. Shi, C. P. Sun, and F. Nori, Controlling
single-photon transport in waveguides with finite cross section,
Phys. Rev. A \textbf{88}, 013836 (2013).

\bibitem{Diaz2015} G. Diaz-Camacho, D. Porras, and J. J. Garcia-Ripoll,
Photon-mediated qubit interactions in one-dimensional discrete and
continuous models. Phys. Rev. A \textbf{91}, 063828 (2015).

\bibitem{Fan2010} S. Fan, S. E. Kocabas, and J.-T. Shen, Input-output formalism for few-photon transport in one-dimensional nanophotonic
waveguides coupled to a qubit, Phys. Rev. A \textbf{82}, 063821
(2010).

\bibitem{Kii2019} A. H. Kiilerich and K. Molmer, Input-Output Theory with
Quantum Pulses. Phys. Rev. Lett. \textbf{123}, 123604 (2019).

\bibitem{Green2015} Ya. S. Greenberg and A. A. Shtygashev,
Non hermitian Hamiltonian approach to the microwave transmission
through a one-dimensional qubit chain. Phys.Rev. A \textbf{92},
063835 (2015).
%
\bibitem{Green2021} Ya. S. Greenberg, A. A. Shtygashev, and A. G. Moiseev,
Waveguide band-gap N-qubit array with a tunable transparency
resonance. Phys.Rev. A \textbf{ 103}, 023508 (2021).
%

\bibitem{Tsoi2008} T. S. Tsoi and C. K. Law,
Quantum interference effects of a single photon interacting with
an atomic chain. Phys. Rev. A \textbf{78}, 063832 (2008).

\bibitem{Chen2011} Y. Chen, M. Wubs, J. Mork, and A. F. Koendrink, Coherent single-photon absorption by
single emitters coupled to one-dimensional nanophotonic
waveguides, New J. Phys. \textbf{13},103010 (2011).

\bibitem{Liao2015} Z. Liao, X.Zeng, S.-Y. Zhu, and M. S. Zubairy, Single-photon
transport through an atomic chain coupled to a one-dimensional
nanophotonic waveguide. Phys. Rev. A\textbf{92}, 023806 (2015).

\bibitem{Liao2016a} Z. Liao, H. Nha, and M. S. Zubairy, Dynamical theory
of single-photon transport in a one-dimensional waveguide coupled
to identical and nonidentical emitters. Phys. Rev. A \textbf{94},
053842 (2016).

\bibitem{Liao2016b} Z. Liao,  X. Zeng, H. Nha, and M. S. Zubairy, Photon transport in a
one-dimensional nanophotonic waveguide QED system Phys. Scr.
\textbf{91}, 063004 (2016).

\bibitem{Zhou2022} C. Zhou, Z. Liao, and M. S. Zubairy,
Decay of a single photon in a cavity with atomic mirrors. Phys.
Rev. A \textbf{105}, 033705 (2022).

\bibitem{Green2023a} Ya. S. Greenberg,  A. G. Moiseev, and A. A. Shtygashev,
Single-photon scattering on a qubit: Space-time structure of the
scattered field. Phys. Rev. A \textbf{107}, 013519 (2023).

\bibitem{Green2023b} Ya. S. Greenberg, A. A. Shtygashev, and A. G. Moiseev,
Time-dependent theory of single-photon scattering from a two-qubit
system. Eur. Phys. J. B \textbf{96}, 162 (2023).

\bibitem{Green2014}  Ya. S. Greenberg,  O. A. Chuikin, A. A. Shtygashev, and A. G. Moiseev,
Dynamical theory of single-photon transport through a qubit chain
coupled to a one- dimensional nanophotonic waveguide. Phys. Scr.
 \textbf{99}, 095119 (2024).

\bibitem{Brown1956} R. H. Brown and R. Twiss, Correlation between photons in two coherent
beams of light. Nature \textbf{177}, 27 (1956).

\bibitem{Zhang2018} Xin H. H. Zhang and H. U. Baranger, Quantum interference and complex photon statistics in waveguide QED. Phys.Rev. A \textbf{97}, 023813 (2018).

\bibitem{Lu2023} Zhi-Guang Lu, Cheng Shang, Ying Wu, and Xin-You
Lu, Analytical approach to higher-order correlation functions in
U(1) symmetric systems.  Phys. Rev. A \textbf{108}, 053703 (2023).

\bibitem{Dom2002} P. Domokos, P. Horak, and H. Ritsch, Quantum description of light-pulse scattering on a single atom in waveguides. Phys. Rev. A \textbf{65}, 033832 (2002).

\bibitem{Lodahl17}
P. Lodahl, S. Mahmoodian, S. Stobbe, A. Rauschenbeutel, P. Schneeweiss, J. Volz, H. Pichler, and P. Zoller, Chiral quantum optics. Nature \textbf{541}, 473 (2017).

\bibitem{RosarioHamman18}
A. Rosario Hamann, C. M¨uller, M. Jerger, M. Zanner, J. Combes, M. Pletyukhov, M. Weides, T. M. Stace, and A. Fedorov, Nonreciprocity realized with quantum nonlinearity. Phys. Rev. Lett. \textbf{121}, 123601 (2018).

\bibitem{Guimond20}
P.-O. Guimond, B. Vermersch, M. L. Juan, A. Sharafiev, G. Kirchmair, and P. Zoller, A unidirectional on-chip photonic interface for superconducting circuits. npj Quantum Information \textbf{6}, 32 (2020).

\bibitem{Blow1990} K. J. Blow, R. Loudon, and S. J. D. Phoenix, Continuum
fields in quantum optics. Phys. Rev. A 42, 4102 (1990).

\bibitem{Gardiner85} C. W. Gardiner and M. J. Collett, Input and output in damped quantum systems: Quantum stochastic differential equations and the master equation, Phys. Rev. A \textbf{31}, 3761 (1985).

\bibitem{Mollow1975} B. R. Mollow, Pure-state analysis of resonant light scattering: Radiative damping, saturation, and
multiphoton effects. Phys. Rev. A \textbf{12}, 1919 (1975).

\bibitem{Glauber1963} R. J. Glauber, The Quantum Theory of Optical
Coherence. Phys. Rev. \textbf{130}, 2529 (1963).

\bibitem{Shen05a} J.-T.
Shen and S. Fan, Coherent photon transport from spontaneous
emission in one-dimensional waveguides, Optics Letters
\textbf{30}, 2001 (2005).

\bibitem{Shen05b} J. T. Shen and S. Fan, Coherent Single Photon Transport in
a One-Dimensional Waveguide Coupled with superconducting Quantum
Bits, Phys. Rev. Lett.\textbf{ 95}, 213001 (2005).

\end{thebibliography}
\end{document}